\documentclass[10pt]{article}

\usepackage{bbm}
\usepackage{bbold}
\usepackage{amsmath}
\usepackage{amsthm}
\usepackage{amssymb} 
\usepackage{mathrsfs}
\usepackage{graphicx}
\usepackage{authblk}
\usepackage[backref=none,hypertexnames=false,colorlinks=true,urlcolor=blue,linkcolor=blue,citecolor=blue]{hyperref}
\usepackage[letterpaper,text={7in,9in},centering]{geometry}
\usepackage{comment}
\usepackage[table,xcdraw]{xcolor}
\usepackage[utf8]{inputenc}
\usepackage{graphicx}
\usepackage{float}
\graphicspath{ {./images/} }
\usepackage{caption}
\usepackage{subcaption}

\usepackage[authoryear,round]{natbib}

\usepackage[english]{babel}

\usepackage{csquotes}

\graphicspath{{eps/}{pdf/}}

\makeatletter\@addtoreset{equation}{section}\makeatother

\begin{document}

\title{The Game is the Game: Dynamic network analysis and shifting roles in criminal networks}
\author[1]{Daniel D.M. Catlin\thanks{ORCID: 0009-0000-9851-4015}}
\author[2]{Giulia Berlusconi\thanks{ORCID: 0000-0003-4654-1059}}
\author[1]{David J.B. Lloyd\thanks{ORCID: 0000-0002-1902-0007}}

\affil[1]{\small School of Mathematics and Physics, University of Surrey, Guildford, GU2 7XH, UK}
\affil[2]{\small School of Social Sciences, University of Surrey, Guildford, GU2 7XH, UK}

\date{}

\maketitle

\vspace{2em}
\noindent \textbf{Corresponding Author:} \\
Daniel D.M Catlin \\
School of Mathematics and Physics \\
University of Surrey \\
Guildford, GU2 7XH, UK \\
Email: d.catlin@surrey.ac.uk

\vspace{2em}
\noindent \textbf{Funding} \\
This work was supported by the Engineering and Physical Sciences Research Council (EPSRC).

\vspace{2em}
\noindent \textbf{Conflicts of Interest} \\
The authors declare that they have no conflicts of interest.

\begin{abstract}
Objectives: This paper explores whether and to what extent individuals' strategic positioning within a criminal network changes over time, and how these changes reflect differential involvement in illicit activities. It incorporates time as a crucial variable to identify key players in criminal networks and assesses the accuracy of the results against the uncertainty around network data collected from criminal justice records.

Methods: Network data are from a judicial document for a two-year investigation targeting a drug trafficking and distribution network. We use Katz centrality in its dynamic version to explore changes in relationships and relative importance of network actors. We then use a novel method of introducing new edges to the network using Bernoulli random trials to simulate missing data and assess the extent to which node rankings based on Katz centrality change or remain the same when introducing some level of uncertainty to our observed network.

Results: We identify actors who consistently held a central role over the course of the two-year investigation and differentiate them from actors who provided key contributions to the group’s activities, but only for a limited period. We show that compared to centrality measures commonly used in criminal network analysis, dynamic Katz centrality is helpful to differentiate individual contributions even among central nodes and explore individual trajectories over time, even when data are incomplete. 

Conclusions: This paper demonstrates the value of key player identification using temporal network data and offers an additional analytical tool to both organised crime scholars trying to capture the complex nature of criminal collaboration and law enforcement agencies aiming at identifying appropriate targets and disrupting criminal groups.

\end{abstract}

\section{Introduction}
\label{introduction}

In the past twenty years, the criminal network perspective has become increasingly popular for the understanding of organised crime \citep{morselli2009}. Its value lies in the combination of theory and methods, where the structure of organised crime groups is not assumed a priori but rather derived from empirical data, and social network analysis becomes a fundamental tool to study how people form, maintain, and sever relations in a context where reciprocal distrust and recourse to violence are common \citep{lampe2009, varese2010, campana2016, bouchard2020}. The term `organised crime' has been used to refer to diverse phenomena over time, with organised crime scholars struggling to agree on a definition \citep{varese2010}. The criminal network perspective aims to ``seek rather than assume" the structure of groups involved in illicit markets \citep[][, p. 18]{morselli2009}, thus accounting for a range of organisational systems, from large hierarchies to small, `flat' criminal enterprises, and allowing to capture the complex nature of criminal collaboration \citep{varese2010, carrington2011}. Criminal actors (or nodes) and their relationships (or ties) become the main focus of the analyses, and techniques from social network analysis provide information on the overall structure of collaboration, patterns of tie formation, and key actors who are well-placed in the network and of potential interest to law enforcement \citep{baker1993, schwartz2009, campana2016}.

Although much progress has been made in relation to the data collection and analysis of criminal networks \citep{bright2022, diviak_method}, several methodological challenges remain. First, the social network analysis of criminal networks is largely cross-sectional and little research exists on the longitudinal evolution and adaptation of criminal organisations to new opportunities or threats, and on changes in individual actors’ strategic positioning and involvement in illicit activities \citep{bouchard2020, bright2021}. Research on criminal network disruption and resilience often focuses on identifying and removing key actors and assessing the impact of such removal on the criminal group's structure and illicit activities, but centrality measures used to identify strategically positioned actors are usually calculated from a single static snapshot of the network based on all available data, and there is a lack of research on the changing importance of particular nodes over time in dynamic settings \citep{carley2006, xu2008, bright2014, bright2017, agreste2016, wood2017, diviak2018, cavallaro2020, toledo2023, taha2024, manzi2024}. Second, while acknowledging the issue of missing data in criminal network analysis, only a few studies consider the impact of missing information on social network analyses and the conclusions drawn from them \citep{campana2012, berlusconi2013, berlusconi2016, bright2022, diviak_method}. The implications of this are both theoretical and practical. The use of static centrality measures on imperfect data can lead to inaccurate conclusions about criminal groups' structures of collaboration while also preventing law enforcement agencies from identifying appropriate targets for monitoring or arrest.

This study addresses both gaps by introducing a dynamic centrality measure to study criminal networks and by conducting a sensitivity analysis to assess the likely impact of missing data on individual nodes' centrality scores and the overall validity of the results. This study uses Katz centrality to analyse temporal network data and assess how actors' positions change over time \citep{katz1953}, with a focus on a drug trafficking and distribution network. We use network data from a two-year investigation conducted by Italian law enforcement agencies targeting a criminal network smuggling cocaine and hashish into Italy. Peculiar to these data is the arrest of a key player mid-investigation, which allows to explore changes in individual positioning in the context of increased law enforcement risk. We contribute to the literature on criminal networks disruption and resilience by showing that dynamic Katz centrality can help differentiate actors based on their contribution to the network over time and identify a small but relevant group of individuals whose consistent involvement in the group's illicit activities makes them of particular interest to law enforcement. We also address the missing data problem by using a novel method of introducing new edges to the network using Bernoulli random trials to simulate missing data and assess the accuracy of our results against the uncertainty around network data collected from criminal justice records. More broadly, this paper contributes to the literature on organised crime and illicit markets by demonstrating how even relatively `flat' and short-lived criminal enterprises are characterised by the presence of a relatively small number of actors who are key in organising and facilitating drug trafficking operations throughout the life-span of the group as well as ensuring its functioning when faced with law enforcement targeting. 

This paper is structured as follows. Section \ref{sec:background} discusses the literature on criminal network disruption and resilience, with a focus on centrality measures commonly used to identify key actors. It also considers two key issues in criminal network analysis, i.e., the dynamic nature of groups operating in illicit settings and the likelihood of missing information on both actors and their relations. Section \ref{sec:methodology} introduces the various mathematical apparatus used to calculate Katz centrality scores and node rankings as well as our novel uncertainty modelling technique. Section \ref{sec:results} presents our results and section \ref{sec:conclusions} includes a discussion of the results and their implications for research and practice.

\section{Background}
\label{sec:background}

A substantial portion of the literature on criminal networks has focused on identifying `key players', that is, actors that are central in the organization and management of the activities of the network and whose removal may lead to network disruption \citep{bright2014, bright2015, bright2015_gc, calderoni2014, calderoni2015, everton2012, carley2006, toth2013, cavallaro2020}. Identifying key actors has several advantages. From the perspective of law enforcement agencies investigating criminal groups, it can help identify suitable targets for monitoring or actors whose removal may lead to the disruption of the network \citep{sparrow1991, bright2021, morselli2010, calderoni2014}. From the perspective of organised crime scholars, it can help identify individuals who are positioned strategically within the network and may have high levels of human and social capital that are worth investigating further \citep{lampe2009, bouchard2020}. It can also provide insights into the organisation of the whole group, i.e., whether it has a centralised, hierarchical structure, a core of closely linked individuals surrounded by a much larger number of peripheral actors, or a `flatter' structure with long communication chains \citep{morselli2008, diviak2018, krajewski2022}. 

There is broad agreement in the literature on the existence of key players in criminal networks, yet less consensus on who they are and whether or how frequently individual network positions shift over time \citep{morselli2010}. Research on organised crime identifies a range of roles within criminal organisations, including (ring) leaders and nodal offenders, coordinators and brokers, low-level suspects, and facilitators \citep{kleemans_depoot, vankoppen2010, morselli2008}. Even where formal hierarchies exist, these do not hold in the context of transit crimes such as drug trafficking \citep{calderoni2012, paoli2004}. Instead, flexibility is the norm, with smaller groups of individuals collaborating on relatively time-limited tasks \citep{kenney2007, bouchard_morselli}. Key players are therefore not necessarily, nor often, high-status individuals fulfilling executive functions who can be described using the kingpin or boss designation. Rather, they tend to be medium-status individuals who organise and coordinate the execution of day-to-day tasks \citep{calderoni2012}. Some degree of stability, however, is required to perform such tasks and to ensure the continuity of criminal networks \citep{bouchard_morselli}. While research points to high turnover in organised crime \citep{kleemans2013}, it remains unclear whether this predominantly affects peripheral actors or extends to more central ones, and more importantly, to what extent individual network positions shift over time, and how \citep{morselli2010}.

Centrality measures are an essential tool in social, and criminal, network analysis. Developed to quantify the structural importance of actors in a network \citep{borgatti2006, freeman1977, freeman1978}, they have become an integral part of the analysis of criminal groups and have been widely used to identify key players within them \citep[see, e.g.,][]{morselli2009}. While centrality measures are key to identify strategically positioned actors within criminal networks, they also present a number of challenges. Although different centrality measures are often correlated with one another \citep{diviak2019}, they operationalise `strategic positioning' in different ways and may identify different individuals as central \citep{morselli2010, bright2015_jccj}. For example, degree centrality is the number of direct connections a person has to others, where higher scores tend to be associated with more access to collaborators and associates, but also with a greater risk of exposure to law enforcement \citep{morselli2010, calderoni2014, xu2004, petersen2011, baker1993}. Betweenness centrality (i.e., the extent to which every actor falls on the shortest path between any two other actors) can instead help identify brokers by locating individuals who facilitate the exchange of information or resources among other network members while simultaneously avoiding too many direct contacts \citep{freeman1977, morselli2008, morselli2010, calderoni2014, grassi2019}. Eigenvector centrality is calculated by finding the leading eigenvector for the corresponding network when it is given in a matrix form and measures the quality of an actor's connections rather than the mere quantity of its connections \citep{bonacich2007}. While eigenvector centrality is not as common as degree and betweenness centralities, it has been used in criminal network analysis to identify actors in leadership positions and predict individuals' sentencing outcomes \citep{calderoni2014, morselli2013}.

While all centrality measures discussed above can help identify key players within criminal networks, they are not without limitations. First, \citet{campana2012} argue for the joint analysis of the structure and the content of criminal connections on the basis that centrality measures can be misleading without additional context from the qualitative information available from wiretap transcripts and other judicial documents \citep{campana2016, varese2013}. For example, \citet{campana2016} demonstrates how prominent actors can sometimes have low degree centrality scores, as in the case of `madams' in a transnational human trafficking network, and warns against interpreting centrality measures without considering additional evidence from court files. Second, the relevance of specific actors can change over time, and yet centrality measures are usually calculated from a single static snapshot of a criminal network \citep{bouchard2020, bright2022}, limiting our understanding of how some actors pause or reduce their involvement in illicit activities over time while others emerge as central. Third, missing data are a common problem in criminal network analysis that can affect centrality calculations and, in turn, our ability to identify key players \citep{campana2012, berlusconi2013}, potentially reducing the effectiveness of law enforcement interventions. The remainder of this section focuses on these last two issues, expands on them, and explains how this paper will address them, thus contributing to advance our understanding of how centrality measures can help us identify key players in the context of imperfect longitudinal criminal network data.

Due to the illicit nature of their activities, criminal organizations often adapt over time to maintain functionality, evade detection, and avoid disruption, while trying to balance the need to pursue their objectives efficiently and the need to avoid law enforcement interventions \citep{morselli2007}. Changes in network structure are unlikely to be planned, but rather emerge from interactions among members of the group \citep{morselli2009}. Scholars have therefore analysed changes in communication and collaboration patterns to explore how the structure of criminal networks evolves over time and how criminal groups respond to external shocks due to law enforcement interventions. For example, terrorist networks become increasingly dense and cohesive as they move from the planning stage to the attack stage \citep{helfstein2011, mcmillan2020}. In contrast, drug trafficking networks tend to become more decentralised over time, especially when targeted by law enforcement agencies \citep{bright2013, bright2019, morselli_petit, duxbury2020}.

Although scholars acknowledge the flexible and adaptive nature of criminal networks \citep[see, e.g.,][]{xu2005, bright2013}, only a handful of studies explore their structural changes over time \citep[e.g.,][]{morselli_petit, bright2013, bright2019, berlusconi2022, diviak2022, diviak2024}. Studies leveraging longitudinal data reveal that networks shift over time in response to external pressures, such as arrests, seizures, and other interventions \citep{berlusconi2022, catanese2016, bright2017, duijn2014}. By constructing a series of temporal snapshots, researchers can track changes in relationships and network structure, identifying periods of increased vulnerability or resilience. This knowledge, in turn, can be used to design effective interventions, predict their outcomes, and prevent reorganisation. Simulations show that small-scale targeted arrests can have a significant impact, but the timing of interventions plays a critical role in their effectiveness, as disrupting the network in its most fragile state can maximize impact and hinder recovery \citep{duxbury2020, manzi2024, yuan2013}.

Studies focusing on criminal network dynamics tend to consider changes in the overall structure of the network and patterns of connections rather than focusing on individual actors and their positioning within the network. Existing research uses exponential random graph models, stochastic actor-oriented models and, more recently, relational hyper-event models, to identify both actor-related and endogenous mechanism behind tie formation \citep[e.g.,][]{berlusconi2022, bright2019, diviak2025}. It emphasises criminal actors' preference for indirect ties and for acting via proxies (i.e., for having a small number of direct ties to highly-connected actors), with direct ties limited to trustworthy associates \citep{bright2019, berlusconi2022, manzi2024}. It does not, however, identify changes in centrality scores for individual nodes. These have only been documented on a few occasions \citep{bright2013, morselli_petit}, as well as high turnover, especially among low-ranking members of the network \citep{diviak2022, bright2019, manzi2025}. Given the focus on key players in the literature on criminal network disruption and the emphasis on the flexible nature of criminal organisations, still very little is known about changes in relationships and relative importance of both core and peripheral actors \citep{bouchard2020}. This is the first contribution that this paper offers.

More specifically, this paper combines key player identification and dynamic network analysis to explore how actors' positions change over time, with a focus on a drug trafficking and distribution network. We use Katz centrality as a centrality measure that can be used with temporal network data, unlike degree, betweenness, and eigenvector centralities \citep{katz1953}. Devised to measure influence in social networks, Katz centrality is used in many fields, from computer science to neuroscience \citep{katz1953, noferini2024, rehm2023}, but it only features in a handful of studies on criminal networks and only in its static version \citep{akartuna2024, taha2024, toledo2023, calderoni2020, cavallaro2020}. The influence of a node within a network is quantified by accounting for both direct connections and indirect connections through neighbours, with longer paths contributing less to actors' individual scores. The relative importance of an actor is therefore higher if they are not only well connected but also connected to other well-connected actors. Katz centrality can therefore help identify actors who find a balance between positioning themselves strategically within the network while favouring indirect relationships with other well-connected network members \citep{calderoni2014, morselli2010}. In its dynamic version, it can help identify actors who consistently hold a central role over time and differentiate them from actors who provide key contributions to the group's activities, but only for a limited period \citep{grindrod2011}. The analysis of shifts in individual positions over time provides insights into criminal networks' coordinating roles -- which are key in ensuring continuity -- without assuming that these are taken by individuals who hold fixed leadership designations, thereby avoiding the conflation of coordinating roles with the `kingpin' or `boss' figure central to traditional hierarchical models of organized crime \citep{morselli2010, bouchard_morselli, calderoni2014}.

When identifying key players in criminal networks, the issue of missing information needs to be addressed as well \citep{budur2015, ficara2021, yeung2025}. Missing data -- in the form of both missing nodes and missing edges -- are a consequence of the very nature of covert networks \citep{diviak_method}. Network members adopt a variety of protection methods to counter electronic surveillance and prevent law enforcement agencies from obtaining evidence, such as regularly searching for bugs in their homes or cars, frequently replacing SIM cards used to communicate with associates, and avoiding telephone communications, especially in the aftermath of a drug seizure or the arrest of other network members \citep{berlusconi2022}. Information on relevant actors and their relations is often extracted from criminal justice records, which only include the evidence collected by law enforcement agencies during their investigations \citep{bright2022}. Furthermore, different types of judicial documents vary in the amount of relational data reported, with wiretap records including a purposive sample of all conversations recorded by law enforcement agencies, and arrest warrants and judgements including an even smaller sample of those conversations \citep{campana2012, berlusconi2013}.

Studies on criminal networks usually acknowledge the likelihood of missing data in the discussion of their limitations, but rarely address it in their analyses \citep{diviak_method}. Research on the impact of missing information on centrality measures suggests that centrality measures are quite robust, especially under small amounts of error, and that their stability is influenced by the type of study and the characteristics of the network \citep{bolland1988, costenbader2003, borgatti2006centrality, xu2008}. Degree and betweenness centrality seem to be more robust where measurement error is not random, such as in the case of criminal networks, where `missingness' is systematic and missing data are more likely to affect peripheral actors than those at the core of the network \citep{berlusconi2013, bright2022}. However, we still know very little about the impact of missing data on criminal network analysis, especially when focusing on network-level measures or when using less common centrality measures. This is the second area our paper contributes to.

In order to do this, this study follows \citet{diviak_method}'s suggestion to not only acknowledge, but also tackle the problem of missing data in criminal networks. First, we calculate Katz centrality scores on the observed network and rank nodes by Katz centrality. We also use qualitative information included in the judicial document to classify nodes based on whether they were under electronic surveillance during the investigation (i.e., whether their phone conversations were wiretapped by the police) or whether they were treated as suspects because of their association with actors in the first group (e.g., because they discussed illegal activities over the phone or they offended together with actors under surveillance). We then introduce new edges to the network probabilistically, taking into account whether the actors were under surveillance by law enforcement during the investigation or whether they were included by association with the first group, and recalculate Katz centrality values and ranking. These new edges represent errors in data collection or cases when this information was successfully concealed from investigators \citep{ficara2021, huisman2009}. Finally, we compare the new ranking of nodes with the original one and assess the impact of the new edges (i.e., possible missing links) on our results, i.e., the extent to which our results change or remain the same when introducing some level of uncertainty to our observed network. This also allows us to highlight the trade-off in temporal network data between accuracy and granularity.

\section{Methodology}
\label{sec:methodology}

\subsection{Data}
The data for this article were extracted from the request for remanding suspects in custody for Operation Cicala, a two-year investigation conducted by Italian law enforcement agencies between November 2008 and July 2010 (612 days). The investigation targeted a criminal network trafficking drugs from Colombia and Morocco to Italy via Spain. Peculiar to this investigation is the arrest of a key player in July 2009 but police monitoring continuing for another year, thereby producing rich network data and allowing to explore changes in individual positioning in the context of increased law enforcement risk. The judicial document contains information on the interactions between the members of the drug trafficking network (i.e., telephone calls and in-person meetings), which was gathered via extensive use of electronic surveillance and other investigative methods (e.g., covert observation). It also contains information on other types of ties shared by the network members (e.g., kinship), the role of each individual in the drug supply chain (i.e., supply, importation, distribution), the tasks they held (e.g., trafficker, support, courier), information on their criminal affiliation (e.g., membership to the 'Ndrangheta), and whether they were under police surveillance. These attributes, together with the qualitative information included in the judicial document, are used to determine the validity of our results. 

The network was first analysed in \citet{berlusconi2022}, which includes further details about the data collection and coding. For the purposes of this study, we focus on records of communication (both via telephone and in person) as proxies of criminal cooperation among the suspects and include time as one of the key variables for our analyses. While the original data were analysed as three binary, undirected matrices covering three separate phases of the investigation, in this article, for each recorded interaction (an edge), we note the persons involved (the nodes that are linked by the edge) and the date it occurs. The time of the interaction is also recorded in the case of telephone communications. This information is used to build both a static representation of the network and, more crucially, a network of time-stamped interactions for our analyses. 

First, we take this network as a single static snapshot of the whole investigation. The resulting network includes 128 persons (or nodes) that share 280 edges, where each edge represents the presence of one or more interactions between any two nodes. The minimum degree is 1 and the maximum is 29 (i.e., network members have communicated with at least one other person during the duration of the investigation, with one person directly communicating with 29 other people). The network is characterised by the presence of a few highly connected nodes and a much larger number of peripheral actors and by a global clustering coefficient of $0.26$, which is typical of small, tightly-knit social networks \citep{ugander2011, morselli2009, calderoni2012}. We then include the date of each communication between any two members of the drug trafficking network to take a series of snapshots over different time frames for our dynamic analysis. These are a series of static networks that display interactions for consecutive periods of 1 day, 7 days and 28 days, respectively.\footnote{The results presented in this article are based on the snapshots of 7 days, but the results were similar for 1 and 28 days and the code to generate them can be provided upon request.} We calculate Katz centrality for both the static and dynamic versions of the Cicala network and compare it to two other centrality measures more common in criminological research, namely, degree and betweenness centrality. All analyses were performed in Python, and the code is available on the corresponding author's GitHub page.\footnote{https://github.com/chikin1993.}$^{,}$\footnote{Unless otherwise stated, parameters were assigned default values chosen to satisfy the theoretical requirements of the corresponding method and to ensure stable numerical performance. Many of these parameters have limited influence on the qualitative behaviour of the results provided they remain within an admissible range. For example, the Katz attenuation parameter $\alpha$ must satisfy $\alpha<\frac{1}{\rho(A)}$, where $\rho(A)$ is the spectral radius of the adjacency matrix. Any sufficiently small value satisfying this condition leads to convergence, and thus a fixed default value of $\alpha=0.01$ was adopted.}

\subsection{Katz Centrality for Static Networks}
\label{sec:Static Katz Centrality}
Katz centrality quantifies the influence of a node within a network by accounting for the total number of walks to all other nodes in the network, attenuated by path length. It incorporates both direct connections and indirect connections through neighbours, with longer paths contributing less to the centrality score. Unlike degree centrality, which considers only immediate ties, Katz centrality assigns higher scores to nodes that are not only well-connected themselves but also connected to other well-connected nodes. Consequently, when two nodes have the same number of direct connections, the one embedded in a more central neighbourhood will receive a higher Katz centrality score than the one connected to other poorly connected nodes (see also \citep{akartuna2024} and \citep{toledo2023} for use and interpretation of Katz centrality in a criminological context). While both Katz and eigenvector centralities account for both direct and indirect connections through walks of all length, the former assigns non-zero scores to all nodes whereas the latter is likely to assign some nodes a score of zero in the case of sparse, disconnected, or directed networks, failing to capture small but potentially relevant differences among nodes in a network \citep{katz1953, newman2010}. This limitation of eigenvector centrality is more pronounced in the case of temporal network data, making Katz centrality a more appropriate choice for the present analysis and for future analyses of directed graphs.

We use static Katz centrality first introduced in \citep{katz1953}, so given the network $G$, we represent this as an $N\times N$ binary adjacency matrix $A$, where $A_{ij}=1$ if there exists an edge between nodes $i$ and $j$ and $0$ otherwise, and $N$ is the total number of nodes in the network. Taking successive powers of the adjacency matrix $A^m$ describes paths of length $m$ between any pair of nodes. The element $A_{ij}^m$, where $m > 0$, thus represents the number of paths of length $m$ between nodes $i$ and $j$. We now define $Q$ as the Katz centrality matrix, which considers the sum of all paths of all lengths that can be traversed on the network, given as the sum of the matrices $A^m$ for all $m\geq0$. Each matrix is also weighted by a factor of $\alpha$, where $\alpha<1$ represents the diminishing importance of successively longer paths. Hence, we express $Q$ as:
\begin{equation}\label{eq:katz_sum}
    Q = I + \alpha A + \alpha^2 A^2 + \alpha^3 A^3 + \ldots.
\end{equation}

The sum given by equation \eqref{eq:katz_sum} is a geometric series, which is convergent if and only if $\alpha < \frac{1}{\rho(A)}$, where $\rho(A)$ is the spectral radius of $A$. Hence, $Q$ can be defined as:
\begin{equation}\label{eq:katz_geo}
    Q = (I - \alpha A)^{-1}.
\end{equation}
The element $Q_{ij}$ of the matrix $Q$ represents the influence of node $i$ to node $j$, so in a static network, we will find that $Q$ is symmetric. To find the centrality value for each node, we sum across the rows of $Q$. This summation of a matrix results in a vector of length $N$, where each element is a numerical value corresponding to the Katz centrality score for that node.

\subsection{Katz Centrality for Dynamic Networks}
\label{sec:Dynamic Katz Centrality}
For dynamic networks, we can use an extension of the static Katz centrality proposed by \citet{grindrod2011} to update Katz centrality over time, as the additional variable can be expressed as a time-ordered series of static networks. A dynamic network $H$ consisting of $K$ time-steps gives $H = \{A_k\}\mbox{ for }k=1,2,\ldots,K$, where $A_k$ gives the adjacency matrix of edges that were added at time-step $k$. Similar to static Katz centrality described in section \ref{sec:Static Katz Centrality}, the $(i,j)$th entry of the matrix product $A_{1} A_{2} \ldots A_{K}$ corresponds to the number of walks from node $i$ to node $j$ of length $m$. To consider walks of all lengths while discounting longer walks by a factor of $\alpha$ for each additional step, we can generalise equation \eqref{eq:katz_geo} for dynamic networks as follows:
\begin{equation}
    Q = (I - \alpha A_1)^{-1}(I - \alpha A_2)^{-1} \ldots (I - \alpha A_K)^{-1},
\end{equation}
which can be expressed concisely as:
\begin{equation}
    Q = \prod_{k=1}^{K} (I - \alpha A_k)^{-1},
\end{equation}
which forms the ordered product of a series of equation \eqref{eq:katz_geo} for each network. As in the static case, $\alpha$ must be chosen to be less than the reciprocal of every matrix in the system, hence we require:
\begin{equation}
    \alpha < \frac{1}{\max(\rho(A_k))},  \qquad\forall k \in \{1,\ldots,K\},
\end{equation}
where $\rho(A_k)$ is the spectral radius \footnote{The spectral radius of a square matrix $A$ is defined as $\rho(A) = \max \{ |\lambda| : \lambda \in \sigma(A) \}$, where $\sigma(A)$ denotes the set of eigenvalues of $A$.} of the matrix $A_k$.

Due to the time-dependant nature of this network, we define the following vectors $b$ and $r$ as the node rankings for nodes that `broadcast' and `receive' information, respectively. The centrality matrix $Q$ will no longer be symmetric and will result in two sets of node rankings for `broadcasters' and `receivers', respectively, depending on whether the final centrality values are calculated by summing across either the rows or columns of the matrix $Q$ with a unit vector $s$. Mathematically, this is given as:
\begin{equation}
    \text{Let }s=(1, \ldots, 1)^T, \text{then } b=Qs, \quad\mbox{and}\quad r=Q^Ts.
\end{equation}

Nodes with high Katz `broadcast' centrality scores are effective at disseminating information throughout the network over time, whereas nodes with high Katz `receive' centrality scores tend to receive more information from other nodes. Here, we define the following vectors $b$ and $r$ as the node rankings for nodes that `broadcast' and `receive' information, respectively. Additionally, we can calculate these vectors iteratively, without the need to compute or store the full matrix $Q$ with the following formulae:
\begin{equation}
    b_0=s ,\qquad b_k=(I - \alpha A_{K-k+1})^{-1}b_{k-1},\qquad \forall k \in \{1,\ldots,K\},
\end{equation}
and
\begin{equation*}
    r_0=s ,\qquad r_k=(I - \alpha A_{k+1})^{-1}r_{k-1},\qquad \forall k \in \{1,\ldots,K\}.
\end{equation*}

\begin{figure}[H]
    \centering
    \includegraphics[width=\linewidth]{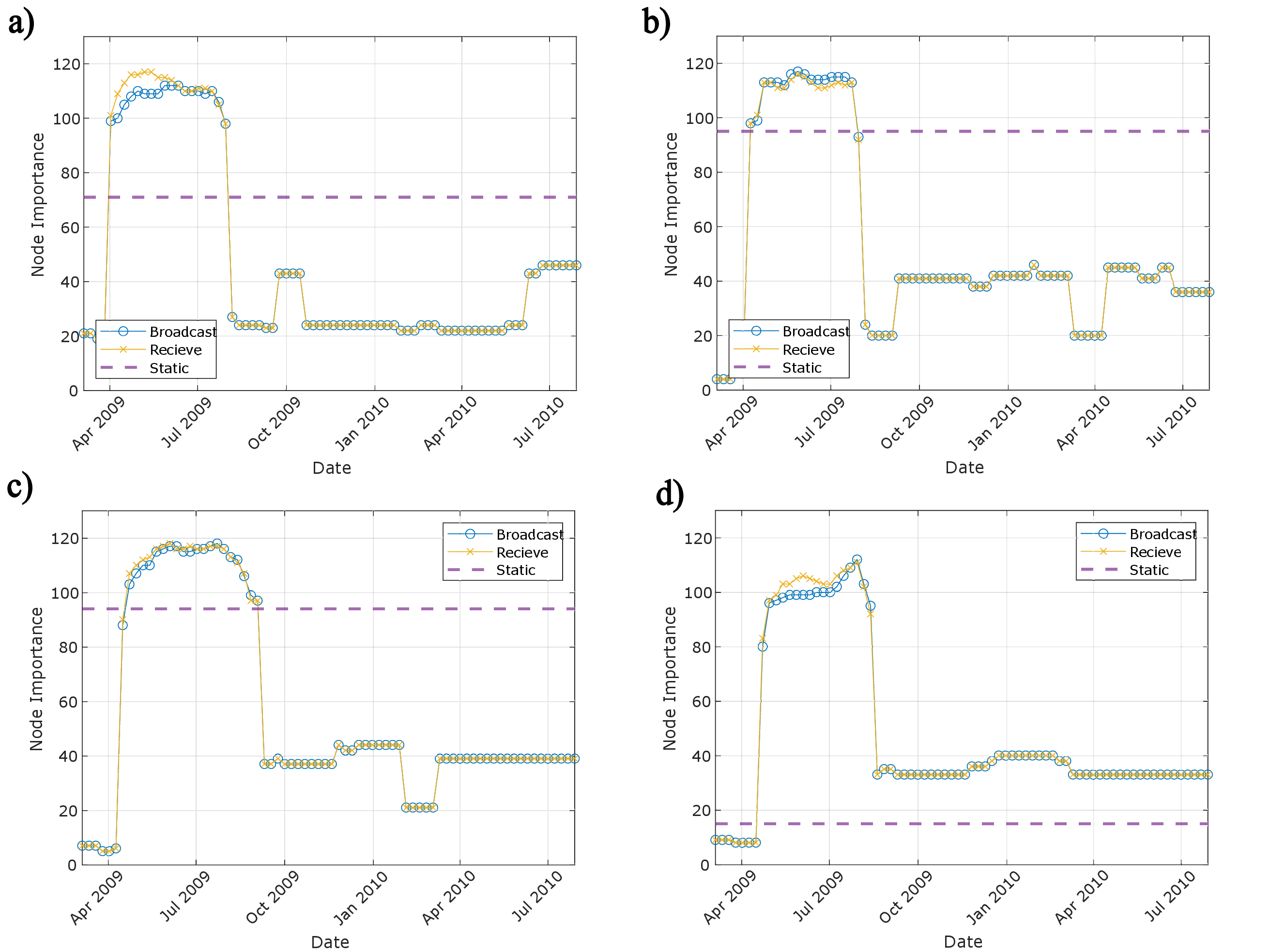}
    \caption{Comparison of receive, broadcast, and static Katz centralities for nodes in the `Withdrawal' cluster, where nodes rank higher at the beginning of the investigation. Panel a) is node 48, b) is node 52, c) is node 55 and d) is node 57.}
    \label{fig:comparison_nodes_absence_old}
\end{figure}

\begin{figure}[H]
    \centering
    \includegraphics[width=\linewidth]{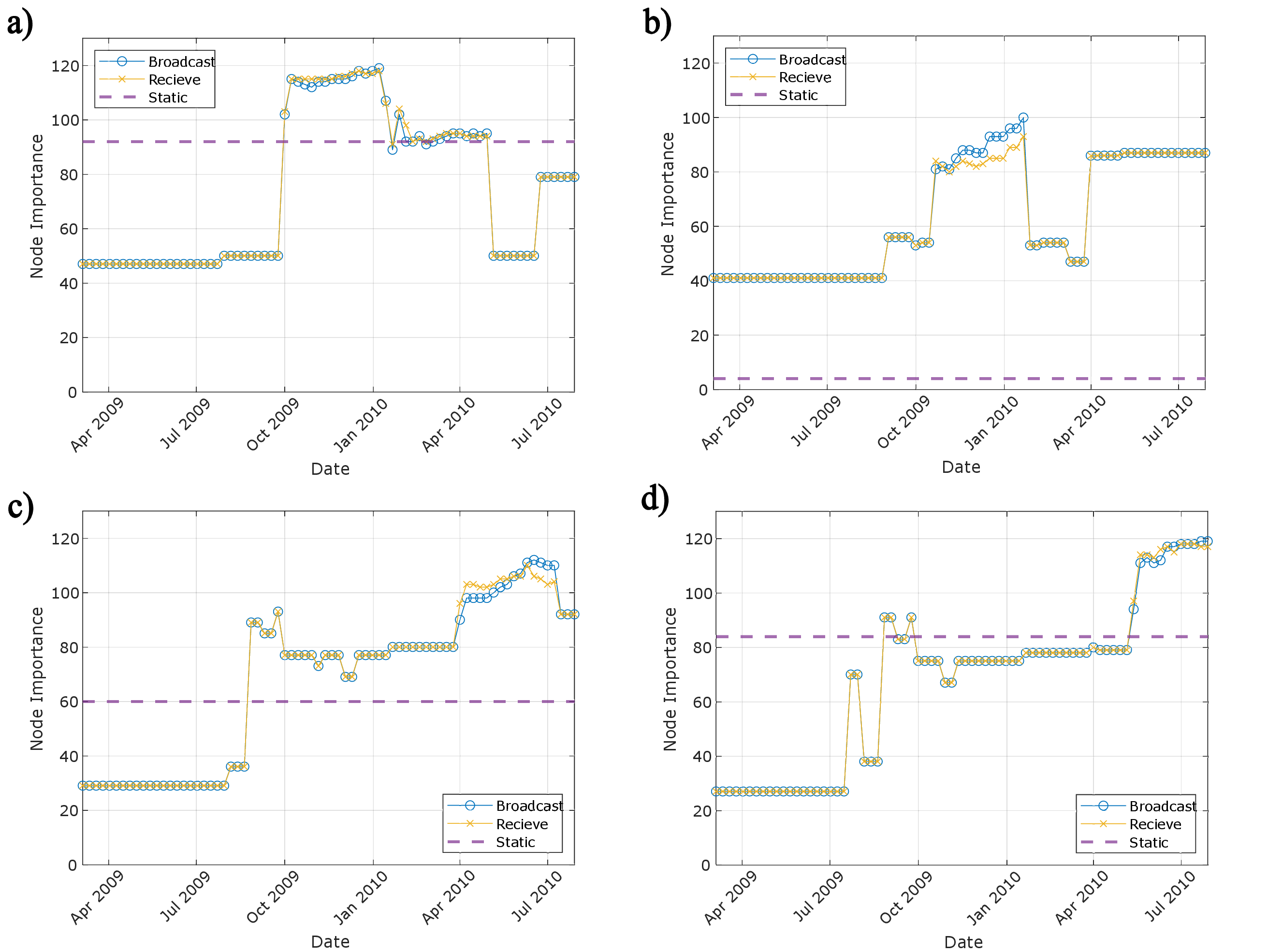}
    \caption{Comparison of receive, broadcast, and static Katz centralities for nodes in the `Emergence' cluster. Panel a) is node 82, b) is node 88, c) is node 100 and d) is node 102.}
    \label{fig:comparison_nodes_fill_old}
\end{figure}

It should be noted that the `broadcast' and `receive' centrality trajectories are very similar for the Cicala network. This suggests that, within this particular dataset, individuals who are effective at disseminating information also tend to be effective at receiving information through those same pathways, leaving little separation between information sources and recipients. Because of these similarities, all reported trajectories for dynamic Katz centrality are drawn from the `broadcast' set only.\footnote{See Figures \ref{fig:comparison_nodes_absence_old}, \ref{fig:comparison_nodes_fill_old}, \ref{fig:comparison_nodes_mangement_old} and \ref{fig:comparison_nodes_mixed_old} in Appendix A for centrality results for `broadcast' and `receive' for the same set of nodes discussed in Section \ref{sec:dynamic_analysis}.} We nonetheless discuss both measures here as they capture distinct aspects of temporal connectivity that are only accessible through dynamic Katz centrality: `broadcast' measures a node's ability to transmit information through time-ordered interactions, while `receive' measures its ability to accumulate information from the network. Although larger or more hierarchically organised criminal networks might exhibit substantial differences between broadcasting and receiving roles, our data cannot confirm whether this convergence is a property of the Cicala network specifically or a general feature of dynamic criminal networks.

\subsection{Katz Centrality with Moving Windows}
\label{sec:moving_windows}

Although the dynamic formulation of Katz centrality proposed by \citet{grindrod2011} can, in principle, be applied to the network as a single time-ordered sequence of snapshots spanning the entire observation period, doing so yields only one centrality score per node, summarising influence across the full 612-day period. To examine how individual positioning evolves throughout the investigation, we instead apply the dynamic Katz centrality formulation separately within each window of a moving window framework, similar to that of a moving average. Each window is itself composed of a time-ordered series of 24-hour static snapshots, so that the temporal structure of interactions within the window is preserved rather than collapsed into a single static network. Repeating this calculation across successive windows produces a time-ordered set of centrality scores for each node, allowing changes in influence to be tracked and compared over the course of the investigation. The complete time series is thus not used in a single calculation, in exchange for the ability to observe how each node's influence develops over time, which is central to the objectives of the present analysis.

The time-stamped communication data for both telephone conversations and meetings are first organised into a series of networks that include all communications over a 24-hour period. Because the number of interactions recorded in the judicial document is fairly small (1,344 interactions over 612 days), there are no communications recorded within many of these 24-hour windows (only 427 24-hour windows out of 612 include any communications). We thus modify the time spans as follows. 

First, we increase the window size, i.e., the number of days that are used to create each network snapshot. Second, we increase the step-size, i.e., the number of days that we move the window forward in time to remove some of the older entries and include newer ones. Both the static and dynamic Katz centrality can be calculated for each of the nodes in the current window and compared with the previous, to produce a time-ordered set of centrality values for each node. Finally, we use a K-means clustering algorithm to group nodes based on their influence by minimising the Euclidean distance between the centres of the clusters and the average position as measured by Katz centrality.\footnote{The Euclidean distance between two points 
$x = (x_1, x_2, \dots, x_n)$ and $y = (y_1, y_2, \dots, y_n)$ in $\mathbb{R}^n$ is 
$d(x,y) = \sqrt{(x_1-y_1)^2 + (x_2-y_2)^2 + \cdots + (x_n-y_n)^2}$.}

We use the K-means clustering algorithm first published in \citet{lloyd1982}. This is an iterative, unsupervised learning algorithm used to partition a set of data (or in our case, nodes) into $K$ distinct clusters based on their common attributes without any other prior information.\footnote{For the initial setup, a random centre for each of the $K$ clusters is chosen, then it proceeds as follows. In the assignment step, each data point (node) is assigned to a cluster by choosing the shortest Euclidean distance from each of the current centre of each cluster and selecting the closest. In the update step, the position of the centre of each cluster is recalculated by taking the mean average of the position of each data point (node) that is assigned to that cluster and set to this value. These two steps then repeat in this order until the centre of each cluster stops changing in the update step due to reaching an optimal convergence and the cluster assigned to each node no longer changes during the assignment step.} A central limitation of K-means clustering is that it requires the number of clusters to be specified a priori and assumes that clusters are approximately spherical and separable in Euclidean space; consequently, the resulting partitions may not adequately capture more complex structures present in the data. Despite this limitation, K-means clustering was deemed appropriate for the present analysis for several reasons. First, the dataset is high-dimensional, with each of the 128 nodes represented by a time series of 74 observations, such that direct visual inspection of the complete dataset would be impractical. An unsupervised learning approach is therefore well suited to identifying underlying patterns and groupings. Second, the relatively modest size of the dataset permits multiple random initialisations and repeated runs across a range of parameter configurations without significant computational cost. Finally, as the objective of the analysis is exploratory (i.e., to identify broad patterns in node behaviour rather than to construct a predictive model), the simplicity, interpretability, and computational efficiency afforded by K-means clustering were considered to outweigh its structural limitations for this purpose.

\subsection{Sensitivity Analysis}
\label{sec:Uncertainty_Quantification}
Networks based on criminal justice data are likely to be incomplete, and missing data are likely to impact on individual nodes' centrality scores \citep{bright2013, bright2022, campana2012, berlusconi2013}. In this section, we introduce a level of uncertainty in our analyses and explore the validity of our results. Our methodology uses a simple machine learning algorithm \citep{wu2012} in conjunction with Katz centrality, following examples where deep reinforcement learning has been used to predict links in dynamic criminal network data with good results when combined with traditional social network analysis \citep{lim2019}.

To account for missing edges that may impact Katz centrality scores, we model the possibility of unobserved interactions between known members of the Cicala network by performing a Bernoulli trial independently for each absent edge. Specifically, for every pair of nodes not connected in the observed network, an edge was added with a fixed probability. This approach provides a simple and transparent mechanism for introducing uncertainty regarding missing interactions while allowing the level of incompleteness to be controlled directly through a single parameter. The maximum number of edges that could be added corresponds to the difference between the number of edges in the observed network and that of the complete graph on the same set of nodes.

We then test how reliable our predictions are to unseen information by comparing the original Katz centrality node rankings to those of the modified network, i.e., by assessing the extent to what our results change or remain the same when introducing some level of uncertainty to our observed network. We start by accepting any edges observed by law enforcement as fact, and by considering every other edge as a connection that might have existed but was not observed. To do this, we introduce new edges to the network probabilistically using individual Bernoulli trials. We then recalculate Katz centrality values and rank nodes by these new centrality values. This allows us to obtain a new ranking of nodes, which can then be compared to the original.

We propose to study the following symmetric, uncertainty adjacency matrix $A_B$, defined as:
\begin{equation}
\label{eq:A_b_substitution}
A_B = A + B (\mathbb{1} - (A+I)),
\end{equation}
where $A$ is the matrix of the observed network of $N$ nodes, $\mathbb{1}$ is the $N\times N$ matrix with all entries being 1, $I$ is the $N\times N$ identity matrix,\footnote{The identity matrix $I$ is an $N \times N$ matrix in which all the elements on the main diagonal (from top left to bottom right) are 1, and all other elements are 0.} and $B$ is a diagonal matrix such that:
\begin{equation}
    B = \mathrm{diag}(B_1, B_2, \ldots, B_N),
\end{equation}
where $B_i \sim \mathrm{Ber}(p_i)$ represents the values of a Bernoulli trial with success probability $p_i$. Thus, $A_B$ is a matrix whose elements are either $1$ where a link is already known (i.e., where a link was observed by law enforcement) or a Bernoulli random variable where the links are not known (i.e., where a link was not observed during the criminal investigation), and zero along the diagonal.

Here we derive a formula to calculate the expectation of the matrix $A_B$ using the matrix $A$ and the uncertainty parameters $p_1$ and $p_2$ to find the expectation of a network with two different levels of uncertainty to model the fact that we are more certain about edges in the network involving nodes who were under surveillance during the investigation as opposed to edges involving nodes who were not under surveillance. We then use qualitative information included in the judicial document to separate the nodes in two groups, i.e., those who were under surveillance by law enforcement during the investigation (65\% of nodes, based on the information from the arrest warrant) and those who were included by association with the first group (i.e., individuals who were treated as suspects because they discussed illegal activities over the phone or offended together with actors in the first group). Any edges that were observed by law enforcement agencies, either via physical or electronic surveillance, are encoded with a weight of 1. The remaining edges are partitioned into the two groups -- under surveillance and not under surveillance. Edges connecting nodes who were not under surveillance are given a weight of $p_1$; edges connecting nodes under surveillance are given a weight of $p_2$. As $p_1,p_2$ are probabilities, we have $p_1,p_2\in[0,1]$ such that when $p_1,p_2=0$, no edges are added and we obtain the observed network. When $p_1,p_2=1$, we obtain the complete graph for this network (i.e., every node is connected to every other node in the network). 

To measure the effect of new edges on Katz centrality scores and rankings, we use the observed network as a baseline for comparison by first setting $p_1,p_2=0$, calculating Katz centrality for each node, and ranking the nodes in order of descending importance. We then simulate networks for each different pair of values of $p_1,p_2$, calculate Katz centrality for each node, and obtain new rankings based on the new centrality scores.\footnote{We start with $p_1$ and $p_2$, but we can generalise this to have a different value for each node, given by: $p_i = p, \forall i \in \{1, \ldots, N\}$ and equations for the expectation of the centrality matrices can similarly be derived, although in this work we restrict ourselves to only using two values.} As we use a Bernoulli distribution, the expectation of these distributions is the parameter $p_1$ or $p_2$. Therefore, equation \eqref{eq:A_b_substitution} can be written as
\begin{equation}
    \mathbb{E}[A_B] = A + (p_1\hat{I}+p_2\tilde{I})(\mathbb{1} - (A+I)).
\end{equation}
Where we require $\hat{I}+\tilde{I}=I$. Furthermore, as the Katz centrality matrix is a manipulation of this, we can derive the expectation of the centrality matrix as a function of $p_1$ and $p_2$ using the definition of Katz centrality to give
\begin{equation}
    \mathbb{E}[Q]=\mathbb{E}[I+\alpha A_B + \alpha^2 A_B^2 + \alpha^3 A_B^3 + \ldots].
\end{equation}
The expectation can then be expanded since expectation is linear and these parts are conditionally independent, therefore
\begin{equation}
    \mathbb{E}[Q]=\mathbb{E}[I]+\mathbb{E}[\alpha A_B] + \mathbb{E}[\alpha^2 A_B^2] + \mathbb{E}[\alpha^3 A_B^3] + \ldots.
\end{equation}
By substitution and rearranging, we see
\begin{equation}
    \mathbb{E}[\alpha A_B] = \alpha (A + (p_1\hat{I}+p_2\tilde{I})(\mathbb{1} - (A+I))).
\end{equation}
As we are using the Bernoulli distribution, we can take the expectation of the variable squared to obtain
\begin{equation}
    \mathbb{E}[\alpha^2 A_B^2] = (\alpha (A + ((p_1\hat{I}+p_2\tilde{I})(\mathbb{1} - (A+I)))))^2.
\end{equation}
It can be seen that these terms are succeeding powers of each other, so they can be expressed as the expectation of the centrality matrix as
\begin{equation}
    \mathbb{E}[Q] = \sum_{n=0}^{\infty} (\alpha (A + ((p_1\hat{I}+p_2\tilde{I})(\mathbb{1} - (A+I)))))^n,
    \label{eq:expectation_series}
\end{equation}
and taking this as a geometric sum in the same way as equation \eqref{eq:katz_geo}, we can calculate this explicitly using
\begin{equation}
    \mathbb{E}[Q] = (I - \alpha(A + (p_1\hat{I}+p_2\tilde{I})(\mathbb{1} - (A+I))))^{-1},
\end{equation}
where we require
\begin{equation}
    \alpha < \frac{1}{\rho(A) + \rho((p_1+p_2)(\mathbb{1} - (A+I)))},
    \label{eq:expectation_bound}
\end{equation}
where $\rho(A)$ represents the spectral radius of a matrix $A$, for the series \eqref{eq:expectation_series} to converge, using the inequality in \citep{weyl1912} to find the bound given in equation \eqref{eq:expectation_bound}.\footnote{Note that each node will have a given probability $p$, which represents the probability that the node will create a new edge. However, new edges between any two nodes will only be created if both nodes pass the Bernoulli trial. In this context, the parameter $q$ such that $q=p^2$ represents the actual probability of a new edge existing.}

To easily compare Katz centrality scores for the nodes in the observed network and in the simulated ones, we convert these values into a ranking system where the node with the highest Katz centrality score has a node importance value of $N$ (where $N$ is the number nodes in the network) and the node with the lowest Katz centrality score has a node importance value of $1$ (for the Cicala network, this corresponds to nodes going from the lowest at 1 to the highest at 128). Translating centrality scores into a ranking system allows us to examine differences in individual node positions across observed and simulated networks. We use a method similar to the Spearman's footrule \citep{spearman2010} and Kendall's tau rule \citep{Kendall_1938} to determine the correlation between two permutations of a set. We find in \citet{Kumar_2010} that these modified formulae are more suitable for our purposes, as they allow us to take into account the importance of each position in the ranking by giving them individual weights. This is a useful feature, as movement of a node to the top of the ranking is more important to us than its movement to the bottom, so we can assign higher weight to the positions at the top of the ranking. 

Given a set of elements $\{1, \ldots, n\}$, let $\sigma$ and $\tau$ be permutations of this set, let $\sigma(i)$ denote the rank of the element i in the permutation $\sigma$, and let $\tau(i)$ denote the rank of the element i in the permutation $\tau$. We then define the footrule distance as $F(\sigma, \tau)$, with respect to our work here, $\sigma$ and $\tau$ both represent the ordered ranking of nodes by importance from the Katz centrality, where $\sigma$ in the original and $\tau$ from the modified version. Next, we must quantify the cost of swapping from one position to an adjacent position, so let $\delta_i$ for $i>1$ be the cost of swapping an element at rank $i-1$ to rank $i$. Furthermore, we now must consider the cost of swapping an element to any other rank, not just an adjacent one, so let $q_1 = 1$ and $q_i=q_{i-1}+\delta_i$ for $1 < i \leq n$. We can now define $\bar{q}_i(\sigma)$ as the average cost of an element moving from rank $i$ to rank $\sigma(i)$ as:
\begin{equation}
    \bar{q}_i(\sigma) = \frac{q_i-q_{\sigma(i)}}{i-\sigma(i)}.
\end{equation}

Hence, we can take the sum across the ranking and generate the formula for position rankings $F_{\delta}(\sigma, \tau)$ as:
\begin{equation}
\label{eq:f_ranking}
    F_{\delta}(\sigma, \tau) = \sum_{i}\bar{q}_i(\sigma)\bigg|\sum_{j:\tau(j)\leq\tau(i)}\bar{q}_j(\sigma) - \sum_{j:\sigma(j)\leq\sigma(i)}\bar{q}_j(\sigma)\bigg|.
\end{equation}

This allows us to give a higher weight to the positions at the top of the ranking, as changes in these reflect the movement of the nodes of highest importance in the network. This is what is used in section \ref{sec:Incorporating_Uncertainty} to compare the node ranking obtained from a modified network with the original ranking obtained from the unmodified, observed network.

It should be noted that Katz centrality is more sensitive to the addition of edges than other centrality measures such as degree centrality. However, when tested, we did not observe any substantive difference between Katz and degree centrality, although much larger and more centralised networks may produce a different outcome. It should also be noted that in this paper we only explore the validity of our results by adding new edges between existing nodes. Adding new nodes would increase the complexity of the calculations and would require specifying the size of the new matrices to account for the additional nodes. Although scholars have highlighted the relevance of `macro networks' \citep{duijn2014} and `opportunity structures' \citep{bouchard_morselli}, the identification and meaningful incorporation of missing nodes remains an empirical challenge, as little is known about actors who operate at the periphery of, or entirely outside, the observed network. We therefore leave the systematic exploration of unobserved nodes to future research. More broadly, our methods are designed with wiretap data and other records of communication in mind. These are generally used as a proxy for criminal collaboration and may not be suitable for the analysis of co-offending, which only includes the joint commission of crimes rather than a wider range of behaviours such as planning, coordination, and strategic communication among network members \citep{calderoni2026}. On a practical level, the computation time for our methods increases exponentially as the network size increases, and the analysis of much larger networks will require more time to compute and more powerful computational resources.

\section{Results}
\label{sec:results}

\subsection{Static Analysis}
We start by examining the use of Katz centrality across the whole set of network data, i.e., all telephone conversations and meetings that occurred between any two actors during the two years of Operation Cicala. We calculate Katz centrality scores for each node in the network and use the K-means clustering algorithm to partition the network actors into two groups. For each node, we also calculate degree and betweenness centrality scores. This analysis allows us to compare Katz centrality to other centrality measures commonly used in criminal network analysis.

\begin{figure}
    \centering
    \includegraphics[width=\linewidth]{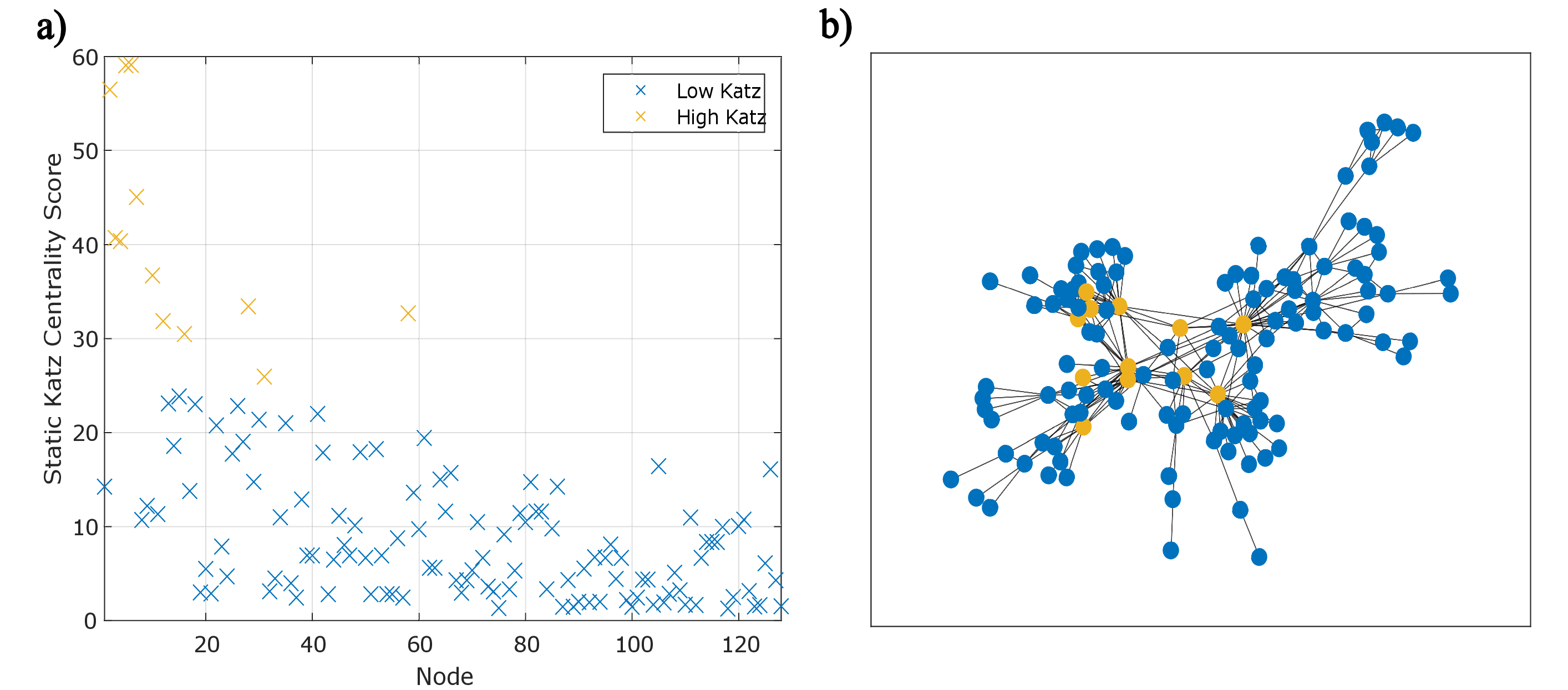}
    \caption{Plot of the static Katz centrality scores for each node in the network when partitioned using K-means clustering. Panel a) shows a plot of the Katz centrality scores for each node, with high-ranking nodes in gold and low-ranking nodes in blue, as identified by the K-means algorithm. Panel b) displays the network diagram for the static network, with edges between nodes representing criminal cooperation and with the same colour groupings overlaid to show the central positions of the high-ranking gold-coloured nodes.}
    \label{fig:comparison_static}
\end{figure}

Figure~\ref{fig:comparison_static}(a) shows that the K-means clustering algorithm partitions the nodes into a smaller group consisting of twelve nodes with high Katz centrality scores, and a bigger group with the remaining of the nodes, who share relatively low centrality scores. Figure~\ref{fig:comparison_static}(b) shows the position of the nodes with high Katz centrality scores in relation to every other actor in the network. The nodes identified in the higher-ranking group are placed centrally in the network and are predominantly known to be traffickers from the qualitative information available in the judicial documents, although other support roles are also present. The presence of a few highly connected individuals and a majority of peripheral nodes is consistent with the findings of other studies \citep{calderoni2014, morselli2010, xu2008}.

\begin{figure}[H]
    \centering
    \includegraphics[width=\linewidth]{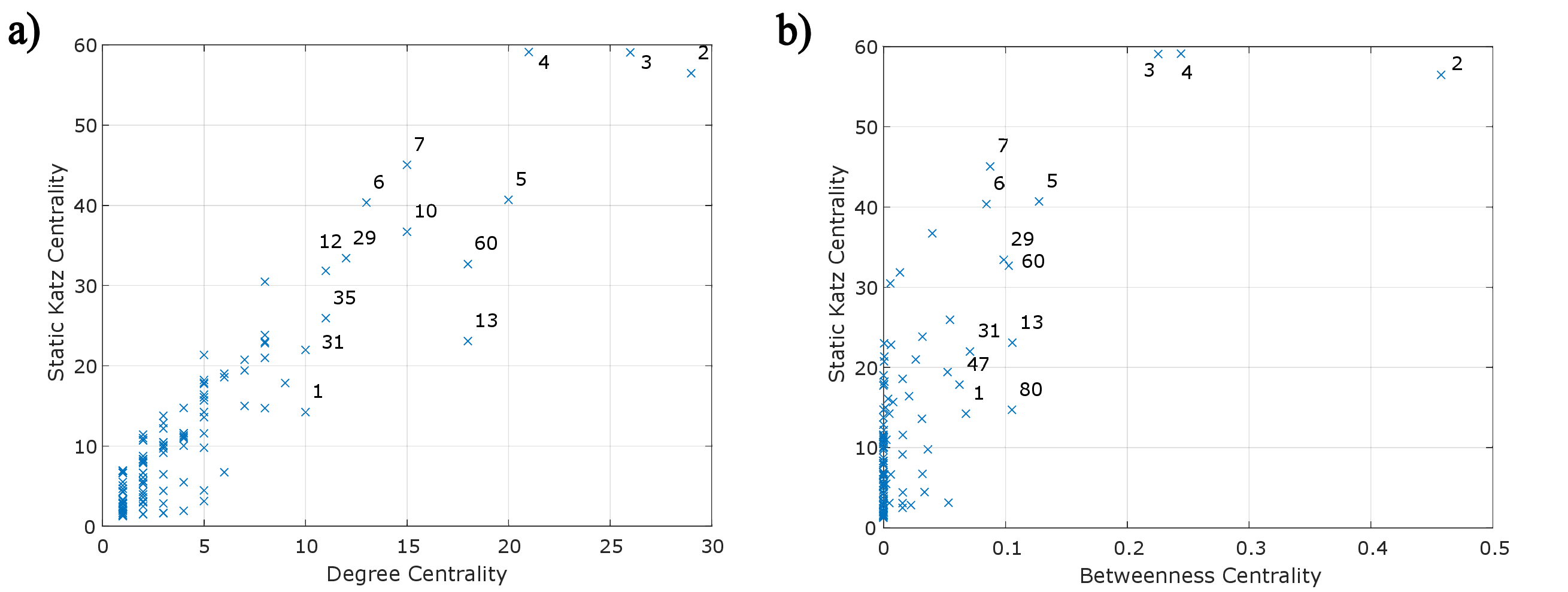}
    \caption{Plots comparing the static Katz centrality scores for each node in the network to degree and betweenness centrality values for the same nodes. Panel a) shows the degree centrality of each node on the x-axis and the static Katz centrality on the y-axis, with node labels for the top 10\% nodes for degree centrality values. Panel b) shows the betweenness centrality of each node on the x-axis and the static Katz centrality on the y-axis, with node labels for the top 10\% nodes for betweenness centrality values.}
    \label{fig:centrality_comparison}
\end{figure}

Figure~\ref{fig:centrality_comparison} compares Katz centrality values with degree and betweenness centrality values. Figure~\ref{fig:centrality_comparison}(a) shows that degree and Katz centralities are strongly correlated, with a Spearman's rank correlation coefficient of 0.78. Network actors with high Katz centrality scores also tend to have high degree centrality scores. Similarly, nodes towards the bottom of the ranking for Katz centrality also tend to rank low for degree centrality. Figure~\ref{fig:centrality_comparison}(b) shows similar results for Katz and betweenness centralities, although the two measures are less correlated, with a Spearman's rank correlation coefficient of 0.54. 

The labelled nodes in the two panels in Figure~\ref{fig:centrality_comparison} correspond to the upper quartile of degree and betweenness centrality scores, respectively. While both Katz and betweenness centrality can help identify nodes in brokerage positions, the latter assigns a value of $0$ to most of the nodes, whereas the former gives them low but non-zero values, allowing identification of small but potentially interesting differences among mid- and low-ranking nodes. For example, node 17 has a betweenness centrality score of $0$ but ranks 52nd (out of 128) for static Katz centrality. In January 2009, he was the courier driving a truck that carried more than five kilograms of cocaine during an attempt to smuggle the drug purchased by node 3 in Spain into Italy. Although only briefly involved in drug trafficking operations, due to his role, he was key in liaising between node 3 (a key player in the drug trafficking network during the first part of the investigation and until his arrest in June 2009), his associates who waited in Milan for the delivery of the drug in vain (nodes 10 and 14), and the recipients of the drug (nodes 11, 49, and 114). Other actors with low betweenness centrality scores are instead much more peripheral overall, and their limited contribution to the activities of the Cicala network is also reflected in very low static Katz centrality scores (e.g., nodes 119 and 120 have betweenness centrality scores of $0$ and rank last for static Katz centrality).

Similarly, Katz centrality builds upon degree centrality by considering both direct and indirect connections between nodes in a network, while also attenuating the influence of longer paths. Therefore, it could help identify actors who may try to position themselves strategically by avoiding a large number of direct contacts and instead communicating with a small number of highly connected individuals \citep{morselli2010}. For example, nodes 122 and 123 have degree centrality scores of $4$ but rank 41st for static Katz centrality. They are brothers and entrepreneurs known by law enforcement agencies for their regular financial support to illegal business ventures by 'Ndrangheta members. Although not some of the most central actors in the network, the qualitative information included in the judicial documents suggests that the two brothers were involved in the financing of the purchase of drugs in exchange for a share and attended meetings with some key traffickers such as node 13 and node 2, who has the highest centrality scores in the Cicala network. The same is true for other buyers, such as nodes 70 and 79 (who both rank 50th for static Katz centrality).

\subsection{Dynamic Analysis}
\label{sec:dynamic_analysis}
We now examine the use of dynamic Katz centrality, discussed in section \ref{sec:Dynamic Katz Centrality}. When calculating dynamic Katz centrality scores for each node, we use a moving window similar to that of a moving average to show how this value changes over time and compare the results to the static Katz centrality ranking. We use a window size of 100 days to establish a long enough time frame for the network to have a sufficient amount of activity during the window period, and a step size of 7 days such that there will be new activity at each time step. The K-means clustering algorithm is used to identify three clusters (which we will call `Coordination', `Emergence' and `Withdrawal' henceforth) and fit each node into one of them. The choice of the number of clusters to identify was trialled from 2 to 4, with the best results obtained with three clusters (see Appendix B). When using two clusters, the results replicate that of the static case, whereas when using four or more clusters, the results become too degenerate due to the relatively small amount of network data. In a purely mathematical sense, the optimal number of clusters could be determined using an elbow plot or another comparable method \citep{thorndike1953}. However, this approach was not considered appropriate in the context of this study, as increasing the number of clusters beyond a small threshold resulted in a substantial loss of interpretability. Due to the relatively small size of the dataset, the clustering procedure could instead be run repeatedly at low computational cost, with the results inspected visually to select an appropriate number of clusters.

\begin{figure}[p]
\centering
\includegraphics[width=\textwidth, height=0.9\textheight, keepaspectratio]{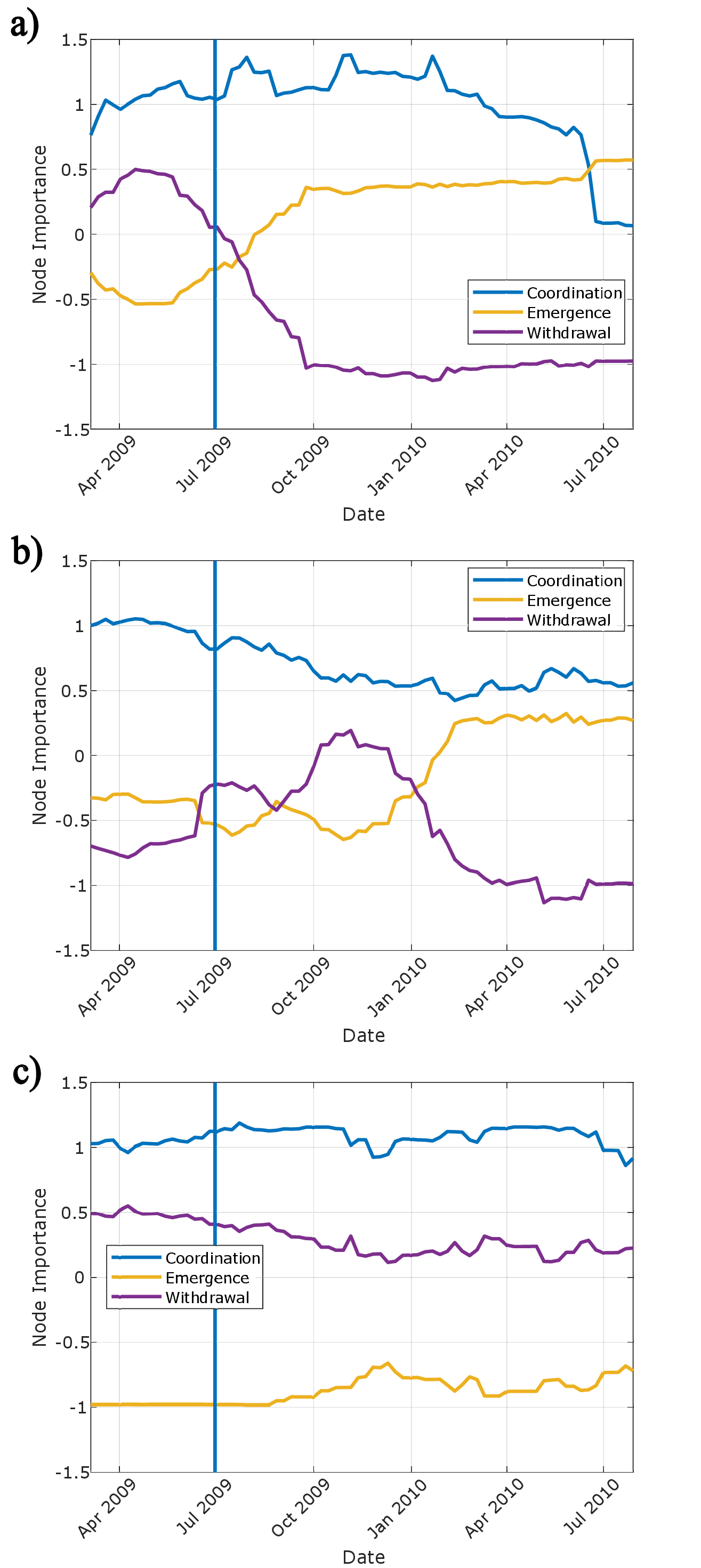}
\caption{K-means centroids for $K=3$ clusters. Panel (a) is for Katz centrality, panel (b) is for degree centrality and panel (c) is for betweenness centrality. The vertical line indicates the arrest of node 3 in June 2009.}
\label{fig:k_means_all_charts}
\end{figure}

We plot the centroid (i.e., the spatial centre) for each cluster at each time frame for each of the three centralities in Figure~\ref{fig:k_means_all_charts}, along with vertical lines indicating the arrest of a prominent drug trafficker made during the investigation. This allows us to compare the relative node importance of actors in each of the three clusters, i.e., how highly nodes in each cluster rank, on average, compared to nodes in other clusters based on their centrality scores. It also allows us to identify any changes in node importance in relation to the arrest by the Italian law enforcement agencies of node 3 -- a key player in the drug trafficking network -- in June 2009, which may be a sign of network re-organisation. While \citet{berlusconi2022} analysed changes in the overall network structure after the arrest, here we focus on changes in individual positions over time, including the emergence of new key actors after the arrest of node 3.

The labelling of the three clusters reflects both patterns of individual positioning of the nodes in each Katz centrality cluster over the course of the two-year investigation and contextual information from the judicial documents about the individuals in each cluster (Figure \ref{fig:k_means_all_charts}(a)). The `Coordination' cluster represents sixteen nodes that have a relatively high importance that is fairly consistent throughout the investigation. While the qualitative information available from wiretap transcripts reveals that four individuals in this cluster had high status within the network, the remaining twelve had instead medium status, suggesting a limited overlap between strategically positioned actors and those with an executive function. Following \citet{vankoppen2010}'s classification, we describe the individuals in this cluster as `coordinators' as they plan and manage concrete criminal activities but are not necessarily in a position of authority. 

Figure~\ref{fig:k_means_all_charts}(a) shows that while individuals in the `Coordination' cluster remained indispensable for the execution of day-to-day tasks and the organisation of drug trafficking activities throughout the investigation, some of them reduced their involvement in the drug trafficking activities towards the end. Although the network was able to adapt and maintain, at least in part, its operational activity after node 3's arrest in June 2009, it suffered from a series of drug seizures that imposed losses to the group and reduced the ability of its members to organise new drug consignments. Lower dynamic Katz centrality scores in the last few months of the investigation may reflect the diminished participation of some key actors actors in drug trafficking activities before law enforcement eventually dismantled the network by arresting most of its members in July 2010.

The `Withdrawal' cluster represents nodes that have a medium node importance score at the start of the investigation, but with significantly lower values from October 2009 onward. This cluster represents mainly people with a medium to low status in the organisation (with notable exception of nodes 3 and 49). They are known to have buyer, courier and support roles, and became less involved in the criminal network's drug trafficking activities after key arrests in mid-2009 are made. While the judicial documents do not offer much qualitative information about the reasons why some of these actors may have left the network or have become less involved in its activities, it is reasonable to assume that some of them may have distanced themselves from the organisation due to fear of arrest or financial losses (e.g., as a consequence of drug seizures) or because they were not involved in new ventures by other network members. For example, node 3’s wife and two sons were also actively involved in the group’s illicit activities but lost prominence after his arrest -- they all belong to the `Withdrawal' cluster.

Finally, the `Emergence' cluster represents nodes whose importance is overall relatively low (except for nodes 80, 82, and 98) but show one or two peaks in rank during the middle of the investigation. These are people who are of less relevance to the organisation at the start of the investigation, but are raised to prominence during the middle of it, after the arrest of one of the network's key players as well as various drug seizures. While, again, limited insights can be drawn from the judicial documents, these nodes are mainly buyers, with a smaller presence of trafficker, support and courier roles, and may have entered the network to replace what was lost during seizures and arrests.

Before we focus on individual nodes within each cluster, we apply the moving window framework and the K-means clustering algorithm introduced in section \ref{sec:moving_windows} to the same data, replacing Katz centrality with degree and betweenness centrality (Figures \ref{fig:k_means_all_charts}(b) and \ref{fig:k_means_all_charts}(c), respectively). In Figure \ref{fig:k_means_all_charts}(b), we still see a cluster representing nodes that have persistent and relatively high importance throughout the investigation along with two other clusters representing individuals who gain and lose prominence over time, respectively. The time when changes in node importance in the `Emergence' and `Withdrawal' clusters become apparent, however, is different for Katz and degree centrality. Katz centrality captures changes in individual positioning earlier than degree centrality, likely because of the truly dynamic nature of this measure. Figure \ref{fig:k_means_all_charts}(c) instead shows that, while nodes in the network can be grouped based on their betweenness centrality values, these remain quite stable throughout the investigation.

Compared to time-windowed degree and betweenness centrality, Katz centrality offers two key advantages. First, it can be applied directly to dynamic networks. When degree or betweenness centrality is used, each time window must be collapsed into a single static network, which inevitably discards temporal information and is particularly problematic when the window spans a longer period. Katz centrality avoids this limitation, as each time window can remain a dynamic network in its own right, thereby preserving the temporal structure of the data throughout the analysis. In addition, although the present study uses undirected networks, Katz centrality extends naturally to directed networks, meaning that the same methodology could be applied without modification should directed data be available. Second, owing to this genuinely dynamic nature, Katz centrality allows changes in individual positioning to be identified earlier than would be possible with other time-windowed centrality measures. Such changes can be analysed both at the aggregate level, by looking at how clusters behave over time, and at the node level, by focusing on specific individuals within the network and how they gain or lose importance over time.

\begin{figure}[H]
    \centering
    \includegraphics[width=\linewidth]{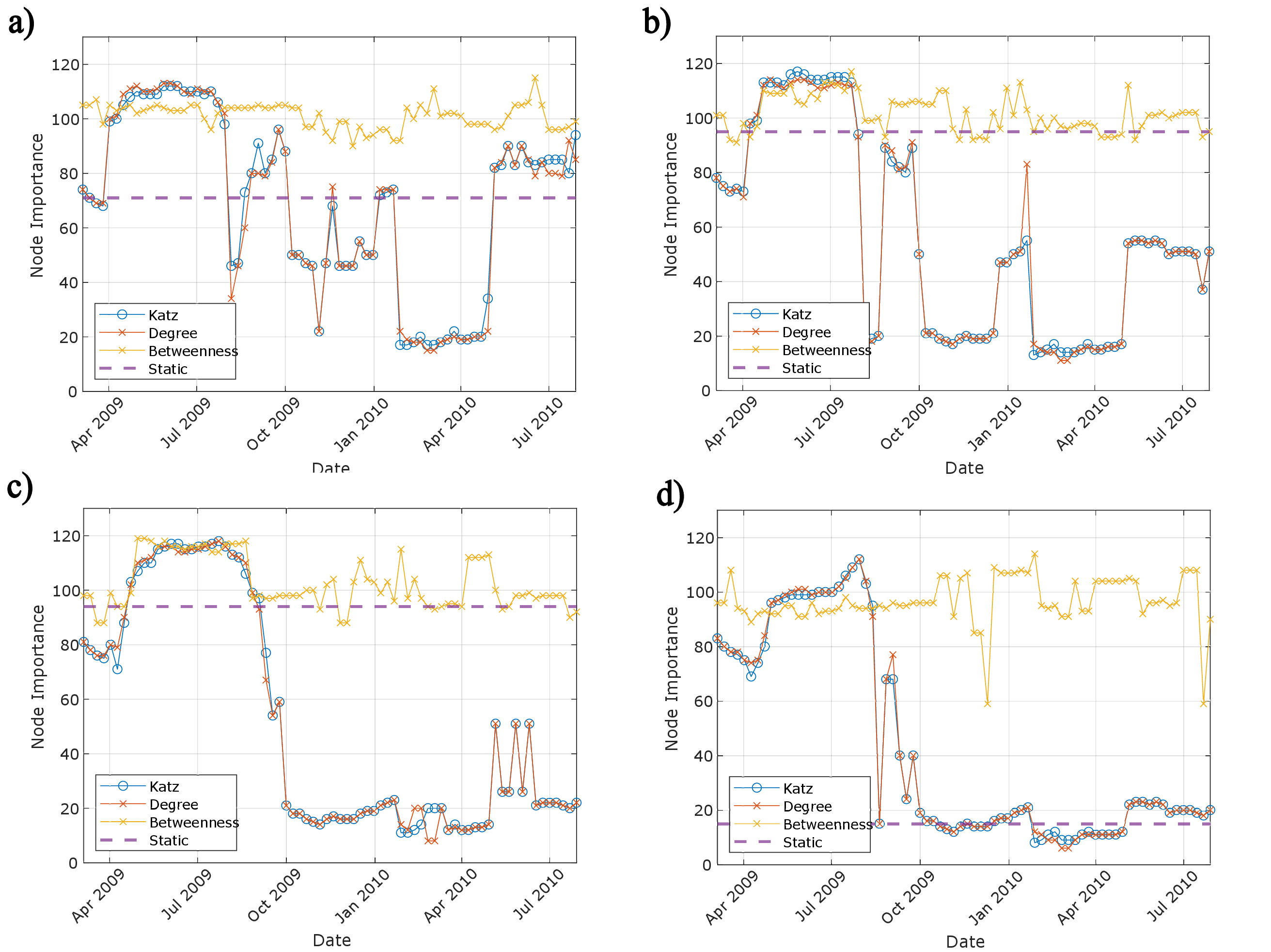}
    \caption{Comparison of static Katz centrality with windowed degree, betweenness and dynamic Katz centralities for nodes in the `Withdrawal' cluster, where nodes rank higher at the beginning of the investigation. Panel a) is node 48, b) is node 52, c) is node 55 and d) is node 57.}
    \label{fig:comparison_nodes_absence_start}
\end{figure}

In Figure~\ref{fig:comparison_nodes_absence_start}(a) and Figure~\ref{fig:comparison_nodes_absence_start}(b), we see examples of the latter, i.e., individuals who lose importance over time and belong to the `Withdrawal' cluster. Nodes 48 and 52 -- who have a support and courier role, respectively -- are  highly placed at the start of the investigation. They were associates of node 3, but both began to lose their places in the network after node 3 was arrested in June 2009. Similarly, nodes 55 and 57 (Figure~\ref{fig:comparison_nodes_absence_start}(c) and Figure~\ref{fig:comparison_nodes_absence_start}(d)) were buyers who would finance the purchase of drugs in producing or transit countries (e.g., Spain) in exchange for a share of the profits from wholesales in Italy. Although the exact reasons for their leaving cannot be fully known, the timing of their reduced involvement in the network coincides with that of drug seizures made by the Italian authorities, suggesting that these losses might have encouraged them to seek out other organisations to conduct their business or to pause or reduce their involvement in drug trafficking activities.

\begin{figure}[H]
    \centering
    \includegraphics[width=\linewidth]{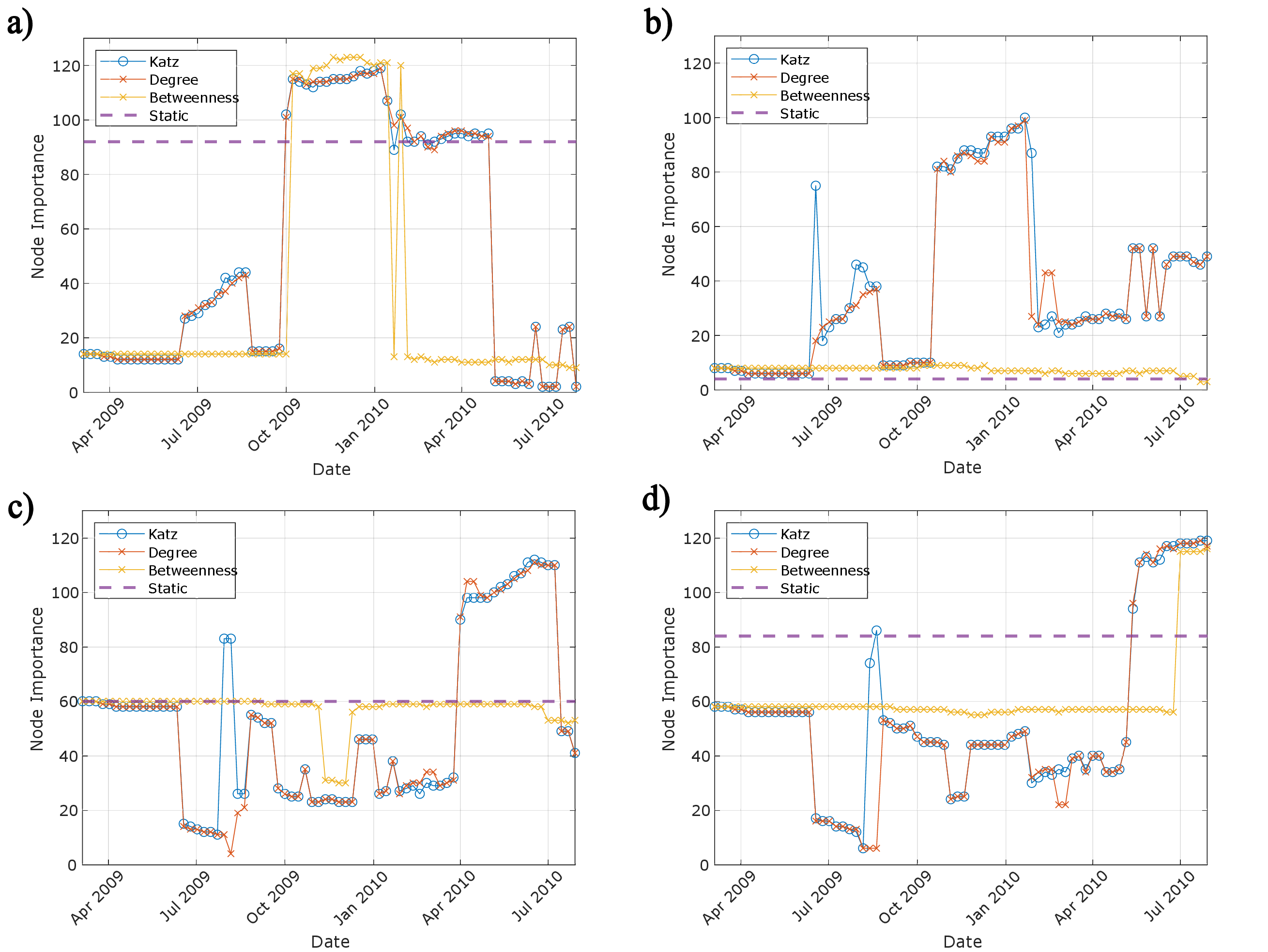}
    \caption{Comparison of static Katz centrality with windowed degree, betweenness and dynamic Katz centralities for nodes in the `Emergence' cluster. Panel a) is node 82, b) is node 88, c) is node 100 and d) is node 102.}
    \label{fig:comparison_nodes_fill}
\end{figure}

Among the actors who belong to the `Emergence' cluster, node 82 -- shown in Figure~\ref{fig:comparison_nodes_fill}(a) -- was a trafficker and the nephew of node 2, who is one of the key traffickers in the network as well as one of those involved in the criminal network activities since the very beginning of the investigation. Beginning in October 2009, node 82 became more involved in drug trafficking activities and began to liaise between nodes 2 and 13 for the purchase and resale of drugs into Italy, filling the gap left by other actors in the `Withdrawal' cluster and increasing his importance during this time. Similarly, nodes 100 and 102 (Figures~\ref{fig:comparison_nodes_fill}(c) and \ref{fig:comparison_nodes_fill}(d), respectively) were associates of node 5, who started as a buyer but later became a supplier of hashish when the network started diversifying into drugs other than cocaine. As node 5 moved into this new role, he received support from other members of his group, including nodes 100 and 102. In Figure~\ref{fig:comparison_nodes_fill}(b) we see node 88, who is a buyer whose involvement in the Cicala network increased after the group suffered losses from a series of drug seizures and tried to involve new buyers to finance new importations to compensate from those losses.

\begin{figure}[htb]
    \centering
    \includegraphics[width=\linewidth]{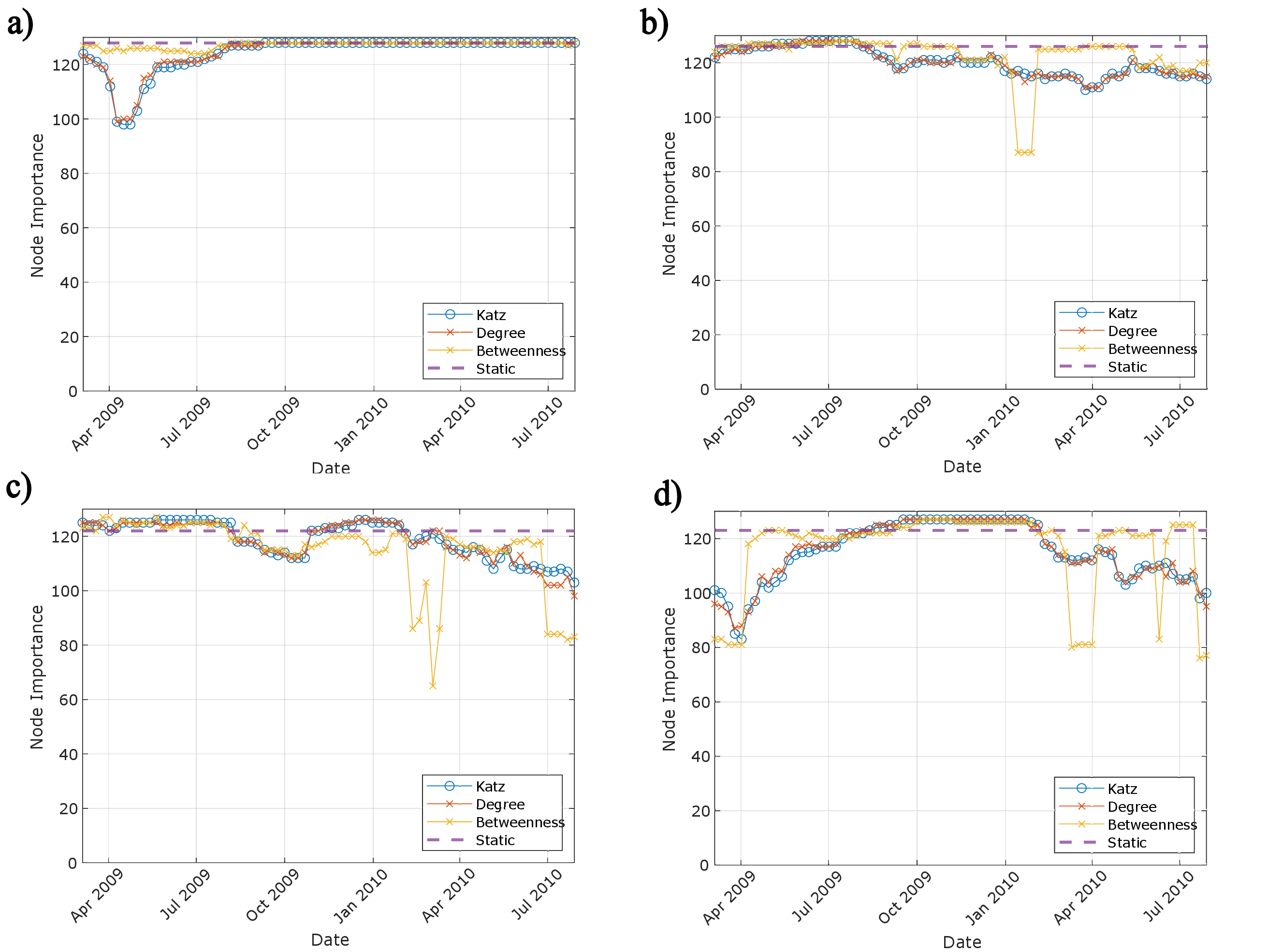}
    \caption{Comparison of static Katz centrality with windowed degree, betweenness and dynamic Katz centralities for nodes in the `Coordination' cluster, where nodes have high rankings throughout the investigation. Panel a) is node 2, b) is node 4, c) is node 7 and d) is node 13.}
    \label{fig:comparison_nodes_mangement}
\end{figure}

\begin{figure}[H]
    \centering
    \includegraphics[width=\linewidth]{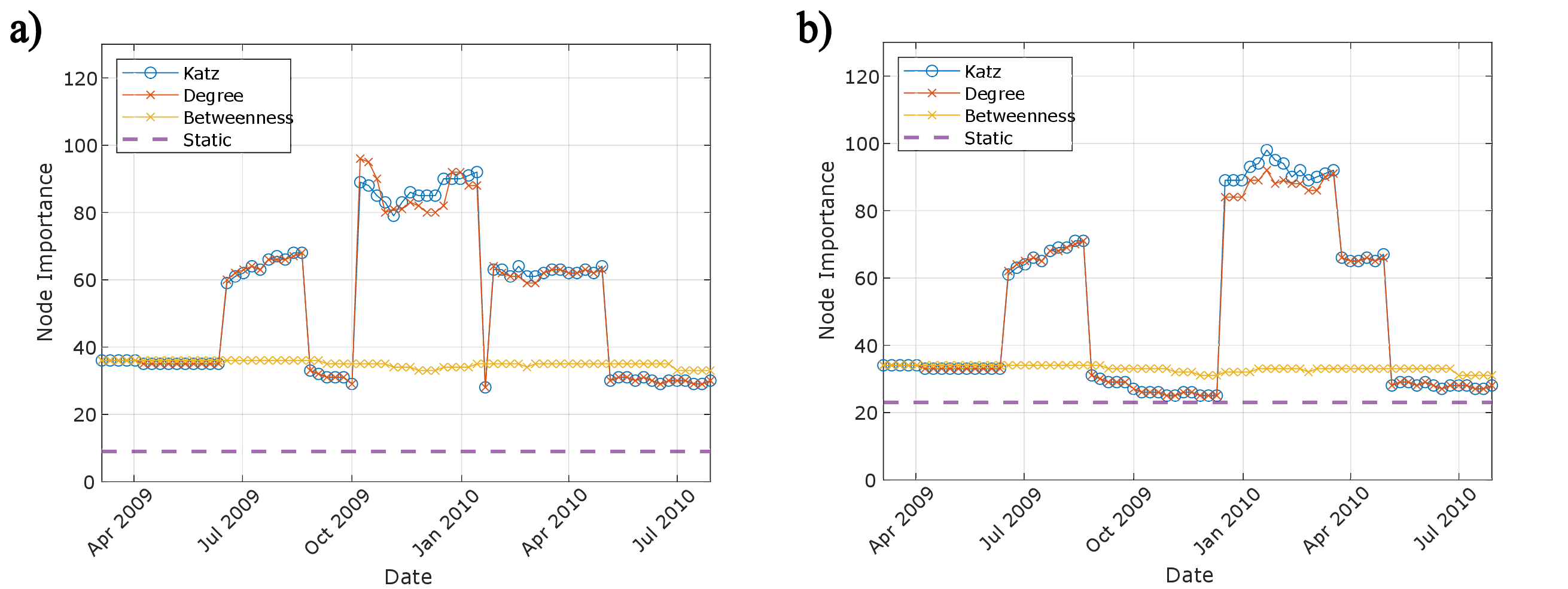}
    \caption{Comparison of static Katz centrality with windowed degree, betweenness and dynamic Katz centralities for two other nodes showing the limits of the purely static ranking. Panel a) is node 124 and b) is 126.}
    \label{fig:comparison_nodes_static}
\end{figure}

Finally, individuals in the `Coordination' cluster show a high node importance that is maintained throughout, although small changes over time are also highlighted. For example, while all nodes in Figure~\ref{fig:comparison_nodes_mangement} had relatively high Katz centrality scores, node 2 (Figure~\ref{fig:comparison_nodes_mangement}(a)) -– a trafficker belonging to the same ‘Ndrangheta family as node 3 –- grew in importance in the first six months of the investigation while node 7 (Figure~\ref{fig:comparison_nodes_mangement}(c)) saw a small decline since July 2009, possibly as a consequence of losing node 3 as one of his main associates. Figure~\ref{fig:comparison_nodes_mangement}(b) shows node 4, who lived in Spain and mediated between Spanish suppliers and Italian traffickers throughout the investigation. Dynamic Katz centrality is also useful to identify actors with low static scores but sporadic, yet consistent, involvement throughout the investigation (Figure~\ref{fig:comparison_nodes_static}). 

\subsection{Sensitivity Analysis}
\label{sec:Incorporating_Uncertainty}
In this section, we introduce a level of uncertainty to the network to account for interactions that may have been missed by law enforcement agencies during the investigation. We incorporate uncertainty in the form of new edges in the network and compare node rankings based on Katz centrality scores before and after the addition of the new edges to assess the reliability of our centrality measures. Our methodology uses a simple machine learning algorithm \citep{wu2012} in conjunction with Katz centrality; a similar approach in combining machine learning with social network analysis techniques can be seen in \citet{lim2019}.

\begin{figure}[p]
    \centering
    \includegraphics[width=\linewidth]{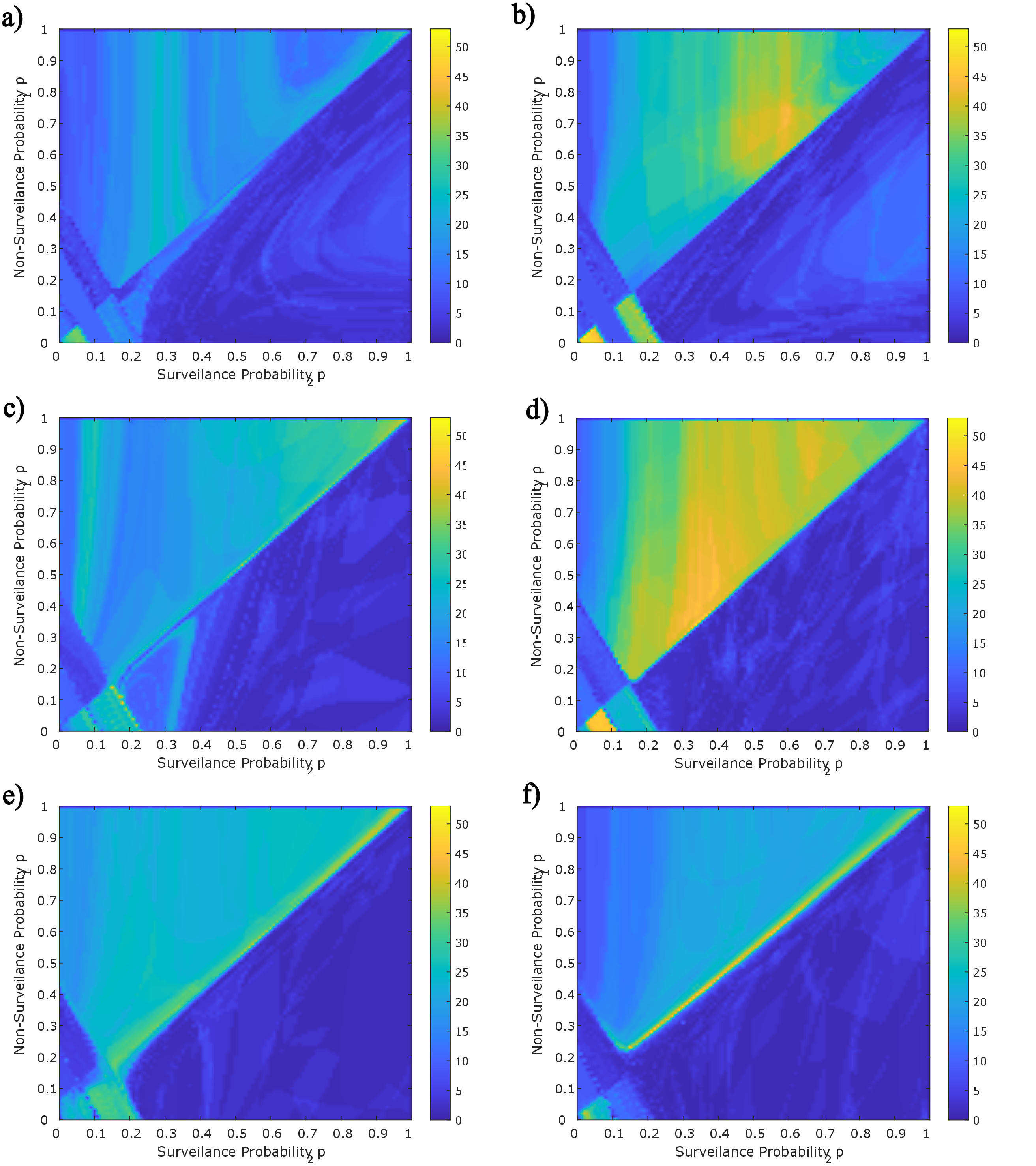}
    \caption{Katz centrality for the dynamic network calculated using varying probabilities $p_1$ and $p_2$ for nodes not under surveillance and under surveillance, respectively, and a step size of $0.01$. Panels a) and b) show the results using 1-day intervals for 'broadcast' and 'receive' Katz centrality, respectively; panels c) and d) show the results using 7-day intervals for 'broadcast' and 'receive' Katz centrality, respectively; and panels e) and f) show the results using 28-day intervals for 'broadcast' and 'receive' Katz centrality, respectively.}
    \label{fig:cicala_surveilence_dynamic}
\end{figure}

Figure~\ref{fig:cicala_surveilence_dynamic} shows the results of uncertainty testing on the Operation Cicala data using the dynamic network.\footnote{As the static Katz centrality matrix is symmetric, summation across either the rows or columns (representing `broadcast' and `receive', respectively) will result in the same centrality vector being obtained and thus no difference between the two in the static case. Figures for the results of uncertainty testing using the static network have been omitted here but can be provided upon request.} Lower (darker) values represent less change in the node rankings after the addition of new edges, and higher (lighter) values indicate that the introduction of new edges has caused significant changes in the node rankings originally calculated on the Cicala data. As we derive formulae introduced in Section \ref{sec:Uncertainty_Quantification} for the expectation of the modified adjacency matrix, all possible unobserved edges are added simultaneously, weighted by the uncertainty parameter $p$. This allows potential new pathways to emerge without the computational overhead of running multiple stochastic trials and averaging the results, preserving computational speed and avoiding any restriction of the method to small networks.

At levels of uncertainty below 0.2 for $p_2$, there is fluctuation in our ranking predictions based on Katz centrality scores. There is also a sharp change when $p_1>p_2$. This is due to the fact that in the static case, we conflated multiple duplicate edges (i.e., multiple interactions) into a single edge and the resulting network contains many more edges and a higher number of short, efficient paths between important nodes, whereas in the dynamic case, each matrix is much more sparse, increasing the probability of each new edge creating a novel route between nodes but only for that time period. This is more pronounced with higher levels of uncertainty for nodes who were not under direct surveillance by law enforcement agencies as there are fewer edges observed between these nodes, and it is more likely that there will be unused paths that are introduced as $p_1$ increases. For nodes under surveillance, there is a higher chance that these more efficient paths are already present, since they were observed by the investigators. Finally, we see areas of higher uncertainty in the results for `receive' Katz centrality rankings compared to `broadcast' rankings, due to the increased uncertainty involved in tracing the flow of information upward through the network, from many receiving nodes back to the fewer originating nodes. The uncertainty compounds when attempting to infer directionality or influence from multiple endpoints, so results derived from this approach should be interpreted with caution.

The comparison of Figures~\ref{fig:cicala_surveilence_dynamic}(a),~\ref{fig:cicala_surveilence_dynamic}(c) and~\ref{fig:cicala_surveilence_dynamic}(e) shows a slight increase in the overall uncertainty of our results as we group together more days of data into longer time frames while still keeping the network in its dynamic form. This shows that the choice of time frame is not an arbitrary one for researchers and may depend on the overall goal of the study and the nature of the network, such as the frequency of important events or member turnover. Lastly, when comparing across Figures~\ref{fig:cicala_surveilence_dynamic}(b),~\ref{fig:cicala_surveilence_dynamic}(d) and~\ref{fig:cicala_surveilence_dynamic}(f), the `receive' results always contain a higher overall level of uncertainty, so more care should be taken when drawing conclusions from these compared to that of the `broadcast' results.

\section{Discussion \& Conclusion}
\label{sec:conclusions}

In this paper, we explored whether and to what extent individuals' strategic positioning within a criminal network changes over time, and did so by addressing two common challenges in criminal network analysis, i.e., the need to incorporate time as a crucial variable to identify key players and the need to assess the potential impact of missing information on individual nodes' centrality scores. We demonstrated that key player identification using temporal network data can improve our understanding of individual actors' strategic positioning and involvement in illicit activities and showed that Katz centrality can be a valid measure to differentiate actors based on their contribution to the activities of a criminal network trafficking cocaine and hashish into Italy, thus offering an additional analytical tool to both organised crime scholars trying to capture the complex nature of criminal collaboration and law enforcement agencies aiming at identifying suitable targets for monitoring and arrest. We found that Katz centrality, along with the K-means unsupervised learning algorithm, can help identify a subset of nodes within the network that are responsible for the coordination of the drug trafficking operations, and can help detect changes in the relative importance of individual actors earlier than other centrality measures. 

In its static version, Katz centrality was highly correlated with degree and, to a lesser extent, betweenness centrality. Static Katz centrality can be useful to identify small but potentially interesting differences among mid- and low-ranking nodes as well as actors who may try to position themselves strategically by avoiding a large number of direct contacts and instead communicating with a small number of highly connected individuals, but it is unlikely to produce a list of key actors that is drastically different compared to one obtained from degree or betweenness centralities. In our case study, the top ten actors identified by static Katz and degree centralities fully overlapped and only differed by one node when comparing static Katz centrality and betweenness centrality (although the node rankings differed in both cases). In its dynamic version, however, Katz centrality revealed its analytical value and highlighted the limits of conflating relational data into a single snapshot of the network to identify central actors.

The dynamic analysis of the Cicala network helped identify actors who consistently held a central role over the course of the two-year investigation and differentiate them from actors who provided key contributions to the group's activities, but only for a limited period. Such differences also exist among the top ten actors identified using static centrality measures. For example, among the top ten players identified using degree and static Katz centralities, only six belong to the `Coordination' cluster, i.e., the group of nodes that have a relatively high node importance that is fairly consistent throughout the investigation. Three of them are part of the `Withdrawal' cluster, suggesting their involvement in the group's activities declined over time, whereas the remaining one belongs to the `Emergence' cluster, i.e., he gained prominence after the arrest of node 3 in June 2009. Similarly, six of the top ten actors for betweenness centrality belong to the `Coordination' cluster, two to the `Withdrawal' cluster, and two to the `Emergence' cluster. In its dynamic form, Katz centrality is thus helpful in distinguishing individual contributions even among central nodes.

Compared to time-windowed degree and betweenness centrality, Katz centrality is also helpful in exploring individual trajectories over time. Paired with a moving window framework, dynamic Katz centrality allows changes in individual positioning
to be identified earlier than would be possible with other time-windowed centrality measures. When applied to the analysis of the Cicala network, this approach enabled us to identify the emergence of new key players soon after the arrest of node 3, thereby suggesting some level of structural reorganisation after law enforcement targeting. Although \citet{berlusconi2022} had already noted changes in the structure of the Cicala network in the aftermath of law enforcement intervention and highlighted ``a reduction in high-status actors’ direct involvement in illicit activities immediately after the arrest of one of the key players and a preference for indirect ties in the final stage of the investigation" (p. 54), our analyses help assess individual actors' shifts in network positions and identify a relatively small number of actors who are key in ensuring the group's functioning when faced with law enforcement targeting.

These results confirm that relations within organised crime groups evolve over time, and that `opportunistic structures' affect collaboration in illicit contexts \citep{varese2010, bouchard_morselli}. As expected from previous literature, the Cicala network was characterised by high turnover and several of its members experienced changes in their positions within the network over time \citep{morselli_petit, bright2014, bright2019, diviak2024, manzi2025}. Although a clear causal relationship between law enforcement interventions and alterations in network structure cannot be drawn, shifts in individual trajectories seem to coincide with law enforcement targeting and especially with the arrest of node 3, a key player in the drug trafficking network during the first part of the investigation and until his arrest in June 2009. This suggests that changes in individual positions are unlikely to be planned but rather emerge from interactions among members of the group as some actors become more cautious and aware of the risk of law enforcement control while others take advantage of new criminal opportunities and emerge as more central \citep{morselli2009}. 

At the same time, however, even relatively flexible partnerships that are fairly horizontal in structure are characterised by the presence of a core group of people who are needed to coordinate the drug trafficking operations throughout the duration of the criminal enterprise. These individuals are able find a balance between positioning themselves strategically within the network while favouring indirect relationships with other well-connected network members \citep{calderoni2014, morselli2010}. They are rarely high-status individuals who are in a position of authority (i.e., what \citet{vankoppen2010} describe as `ring leaders' or `nodal offenders'). Rather, they tend to be medium-status individuals who plan, organise and manage concrete criminal activities such as acquiring the funds to purchase drugs, organising the purchase, smuggling, and wholesale distribution of drug shipments, and ensuring the group's operational safety (e.g., settling disputes between group members, managing communications) \citep{calderoni2012}. These findings confirm that although key players are needed to ensure some level of continuity of a criminal network, hierarchies do not necessarily hold in the context of drug trafficking, such that those in coordinating roles rarely correspond to the traditional boss or kingpin figure \citep{paoli2004, bouchard_morselli}.

Finally, we used a novel method of introducing new edges to the network using Bernoulli random trials to simulate missing data and assess the accuracy of our results. Rankings of nodes based on static Katz centrality were fairly stable until about $20\%$ of the network is altered. In the case of dynamic Katz centrality, we found higher levels of uncertainty for nodes who were not under surveillance during Operation Cicala and when grouping together more days of data into longer time frames, indicating that when using dynamic network data, it is important to consider the trade-off between accuracy in our predictions and granularity of the data used, and how the chosen time frame may affect this. Although it remains impossible to fully appreciate the extent of missing data in criminal network analysis, our results suggest that information about the source of relational data and the scope and degree of law enforcement surveillance (e.g., the proportion of network actors who were under direct surveillance) could help assess the accuracy of the results.

In this study, we relied on network data and other qualitative information retrieved from judicial documents made available by Italian law enforcement and criminal prosecution agencies and used records of communication among suspects as proxies of criminal cooperation. The Cicala network was drawn from judicial documents similar to those commonly used for the network analysis of criminal organisations \citep{bright2022, diviak_method}. The network itself also shares several characteristics that are similar to those of other drug trafficking networks analysed in the literature, such as low density and the presence of a small number of highly connected individuals and a majority of peripheral nodes \citep[e.g.,][]{calderoni2014, morselli2010, xu2008}. While the application of dynamic Katz centrality to other criminal networks may reveal additional insights into the strategic positioning of individual actors within networks over time, it seems reasonable to expect results that are both in line with those of the current study and with the literature on organised crime and criminal network disruption and resilience \citep[e.g.,][]{morselli2009, bouchard2020, bouchard_morselli}. However, despite some promising findings, a wider empirical base is needed to assess the strength of our method.

Another advantage of Katz centrality is that it can be used on directed networks, i.e., networks where ties represent the flow of information or resources, and where senders and receivers can be clearly identified. For example, it is suitable to study networks representing money laundering actions and schemes \citep{akartuna2024} or networks derived exclusively from records of telephone conversations wiretapped by law enforcement agencies \citep{berlusconi2016}. Future work could fit into two categories. First, it could apply the same analytical techniques used in this paper to other criminal networks, including directed networks, to verify whether our results are an exception or to be expected for most criminal organisations and validate the outcomes of this study. Second, it could extend this work to use multilayer networks (based, e.g., on different types of roles or ties that the same set of actors shares) to identify more efficient disruption strategies by removing nodes that may appear of relatively low importance in the overall network but of high importance in individual layers.

The findings generated from this study are of great significance, both academically and practically. From a research perspective, the dynamic analysis of individual positions allows to capture a relevant yet under researched aspect of the `flexible order' that characterises criminal organisations, where ``[p]ositioning and remaining flexible is the key'' \citep[][, p. 11]{morselli2009} and central actors emerge (and fade into the background) within the structure of interactions among those involved in the criminal enterprise rather than being the result of careful planning or complex hierarchies. From a law enforcement perspective, understanding how individuals adapt and shift between different responsibilities and functions within a dynamic environment can help with the identification and targeting of key players, especially in the context of non-hierarchical networks where arresting a single actor is likely to have limited disruptive effect \citep{bright2017, bright2021}. In this context, dynamic Katz centrality could help identify both actors on an ascending trajectory -- who may be targeted to cause the most disruption to the group's activities -- and descending actors, who may be targeted for intelligence gathering purposes (e.g., to convince them to become police informants).

\bibliographystyle{apalike}
\bibliography{references}

\section{Appendix A}
\label{sec:appendix}
\begin{figure}[H]
    \centering
    \includegraphics[width=\linewidth]{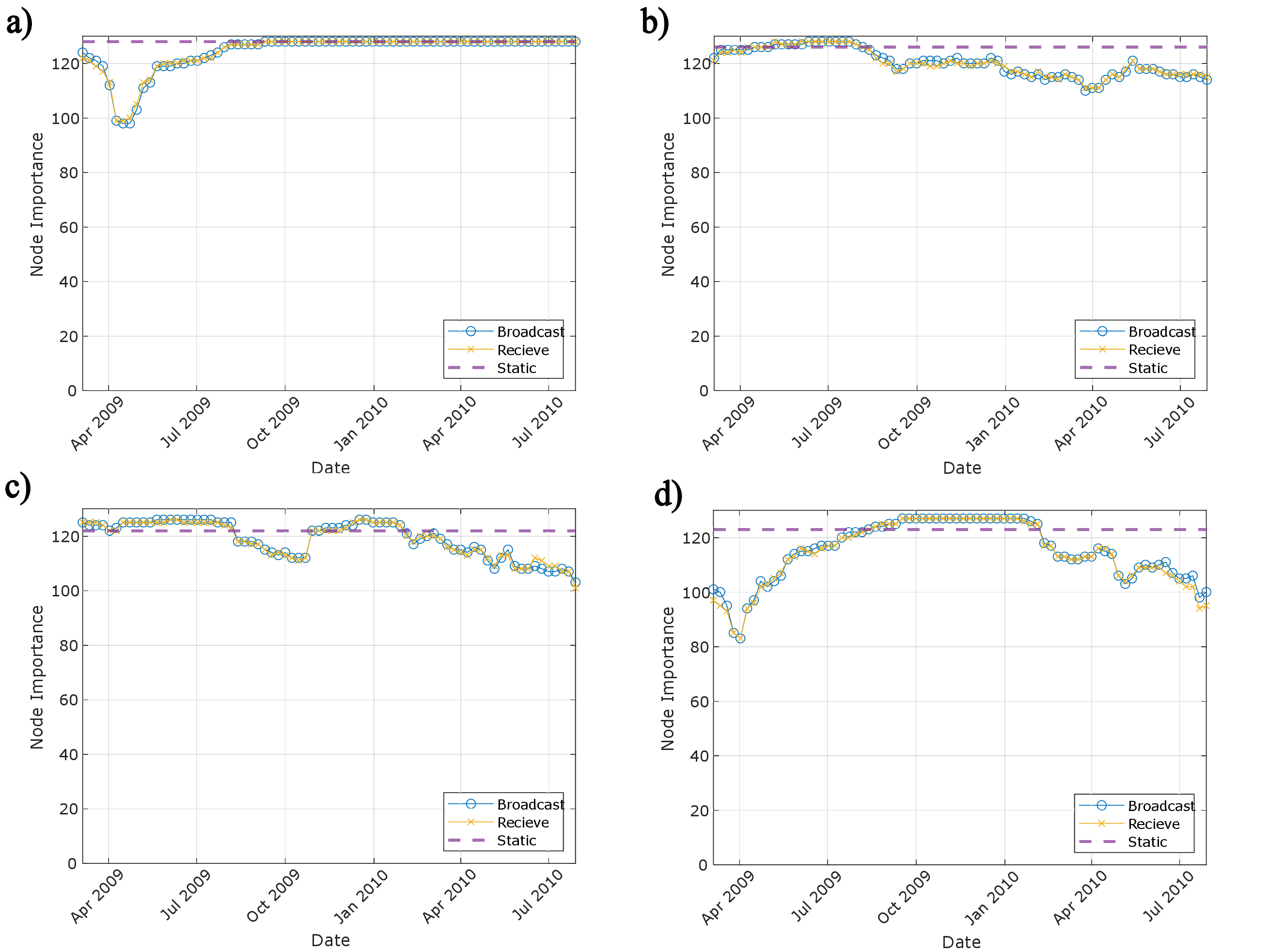}
    \caption{Comparison of receive, broadcast, and static Katz centralities for nodes in the `Coordination' cluster, where nodes have high rankings throughout the investigation. Panel a) is node 2, b) is node 4, c) is node 7 and d) is node 13.}
    \label{fig:comparison_nodes_mangement_old}
\end{figure}
\begin{figure}[H]
    \centering
    \includegraphics[width=\linewidth]{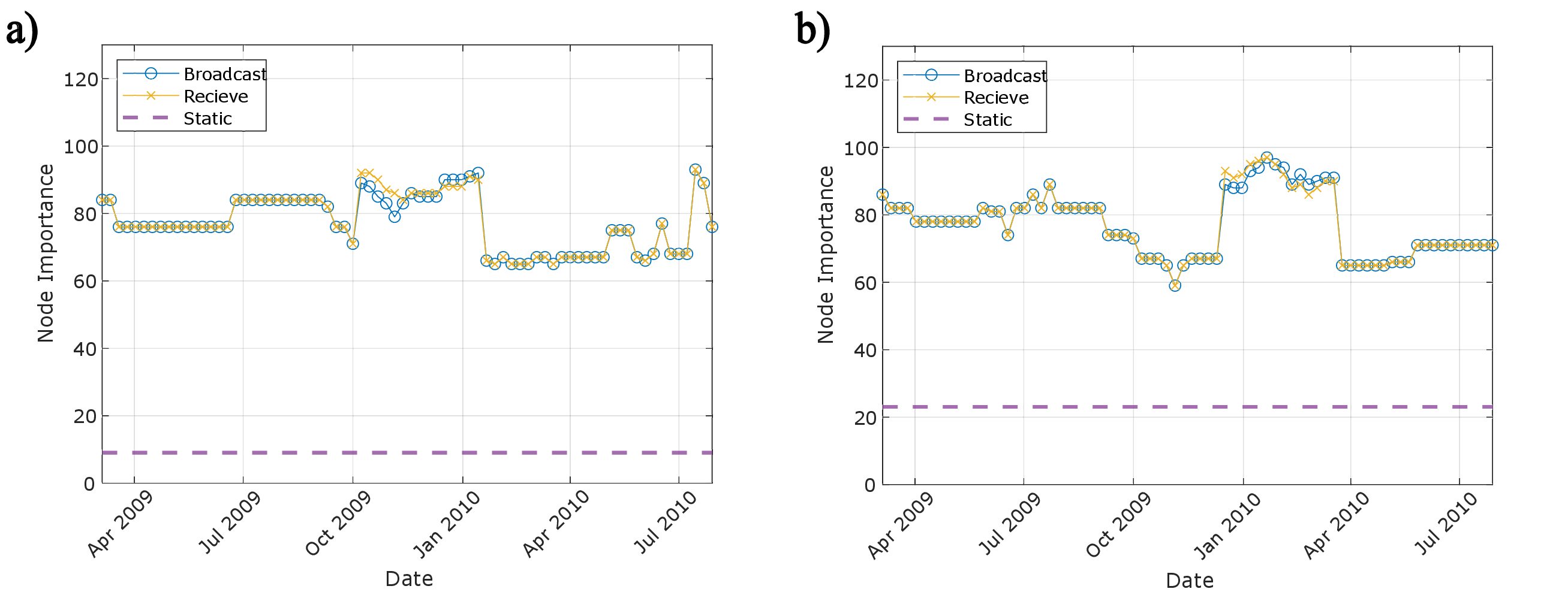}
    \caption{Comparison of receive, broadcast, and static Katz centralities for two other nodes showing the limits of the purely static ranking. Panel a) is node 124 and b) is 126.}
    \label{fig:comparison_nodes_mixed_old}
\end{figure}

\section{Appendix B}
\begin{figure}[p]
    \centering
    \includegraphics[angle=90,width=0.9\linewidth]{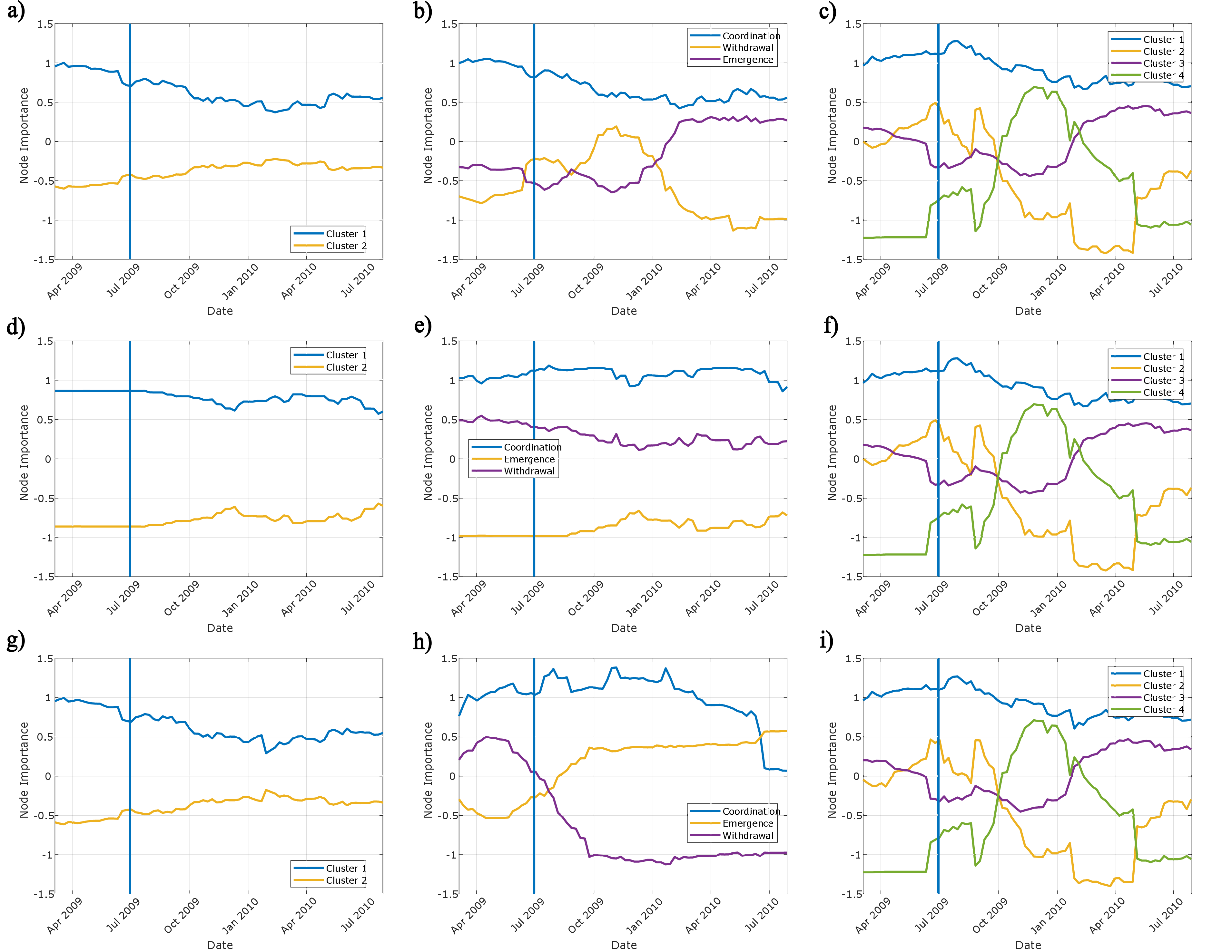}
    \caption{K-means cluster centres for the degree, betweenness and Katz centralities, split into $K\in\{2,3,4\}$ clusters, with the blue horizontal line indicating the arrest of node 3. Panels are as follows: a) Degree, $K=2$, b) Degree, $K=3$, c) Degree, $K=4$, d) Betweenness, $K=2$, e) Betweenness, $K=3$, f) Betweenness, $K=4$,  g) Katz, $K=2$,  h) Katz, $K=3$,  i) Katz, $K=4$}
    \label{fig:cicala_clustering_3x3_grid_named}
\end{figure}

\end{document}